\documentclass[14pt,useAMS,a4paper,usenatbib]{emulateapj}
\usepackage{epsfig}
\usepackage{amsmath}
\usepackage{natbib}

\begin{document}

\title{Methyl Acetate and its singly deuterated isotopomers in the interstellar medium}

\author{Ankan Das\altaffilmark{1}, Liton Majumdar\altaffilmark{2,3,1}, Dipen Sahu\altaffilmark{1}, 
Prasanta Gorai\altaffilmark{1}, B Sivaraman\altaffilmark{4}, Sandip K. Chakrabarti\altaffilmark{5,1}}

\affil{$^1$Indian Centre for Space Physics, Chalantika 43, Garia Station Rd., Kolkata, 700084, India.}
\affil{$^2$Univ. Bordeaux, LAB, UMR 5804, F-33270 Floirac, France.}
\affil{$^3$CNRS, LAB, UMR 5804, F-33270 Floirac, France.}
\affil{$^4$Space and Atmospheric Sciences Division, Physical Research Laboratory, Ahmedabad, 380009, India.}
\affil{$^5$S. N. Bose National Centre for Basic Sciences, Salt Lake, Kolkata, 700098, India.}

\email{$^1$ankan.das@gmail.com}

\shorttitle{Formation of Methyl Acetate in the ISM}
\shortauthors{Das et al.}

\begin{abstract}
Methyl acetate ($\mathrm{CH_3COOCH_3}$) has been recently observed by IRAM $30 \ \mathrm{m}$ radio telescope
in Orion though the presence of its deuterated isotopomers is yet to be confirmed.
We therefore study the properties of various forms of methyl acetate, namely, $\mathrm{CH_3COOCH_3}$, 
$\mathrm{CH_2DCOOCH_3}$ and $\mathrm{CH_3COOCH_2D}$. Our simulation reveals that these species could be
produced efficiently both in gas as well as in ice phases. Production of methyl
acetate could follow radical-radical reaction between acetyl ($\mathrm{CH_3CO}$) and methoxy 
($\mathrm{CH_3O}$) radicals. To predict abundances of $\mathrm{CH_3COOCH_3}$ along with its two
singly deuterated isotopomers and its two isomers (ethyl formate and hydroxyacetone), 
we prepare a gas-grain chemical network to study chemical evolution
of these molecules. Since gas phase rate coefficients for methyl 
acetate and its related species were unknown, either we consider similar rate
coefficients for similar types of reactions (by following existing data bases) or we carry out 
quantum chemical calculations to estimate the unknown rate coefficients. 
For the surface reactions, we use adsorption energies of 
reactants from some earlier studies. Moreover, we perform quantum chemical calculations to obtain
spectral properties of methyl acetate in infrared and sub-millimeter regions. 
We prepare two catalog files for the rotational transitions 
of $\mathrm{CH_2DCOOCH_3}$ and $\mathrm{CH_3COOCH_2D}$ 
in JPL format, which could be useful for their detection in 
regions of interstellar media where $\mathrm{CH_3COOCH_3}$ has already been observed.
\end{abstract}


\section{Introduction}

According to the Cologne Database for Molecular Spectroscopy \citep{mull01, mull05}, 
about $180$ molecules have been observed in the interstellar medium (ISM) or in 
circumstellar shells, among which several species are organic in nature. Several bio-molecules 
(amino acids) have been observed in the meteorites discovered on Earth which indicate that 
such complex bio-molecules must be produced in frigid and tenuous interstellar space. In fact it is long believed that 
basic biomolecules on earth could have been brought through numerous meteoritic bombardments in the early stage of earth's life
\citep{oro61,oro79,dels81}.
The sea water which is also produced by such bombardments (see, Sarafian et al. 2014 and references therein) should also 
carry complex pre-biotic species which in presence of tidal forces of the sun and the moon could have 
helped formation of even more complex molecules on sea surface \citep{chak12}.

Quantitative computation of the abundances of complex pre-biotic molecules in collapsing clouds
and star forming regions have started only recently
\citep{chak00a,chak00b,chak15}. It is suggested that even if the abundance of adenine is small enough for detection with
present day technology, its immediate precursors should be observed in interstellar space \citep{maju13}.
Depending on physical conditions, these precursor molecules could be formed by various gas-grain interaction processes.
Several works have been done in the past to model chemistry inside both the diffuse and dense 
molecular clouds \citep{vand86,fede95,fede96,lepe04}.  
A large number of models have been developed over the last few years 
\citep{hase92,das13a,das13b,das15,maju14a,maju14b,das10,das11,chak06a,chak06b,sahu15}
to explain chemical composition of interstellar gas and grains under various physical circumstances.

Methyl acetate is a flammable substance and because of the low toxicity and fast evaporation rate, 
it is occasionally used as a solvent \citep{sipo08}. It is the simplest easter and thought to be 
a potential prebiotic candidate which could be synthesized   
in the ISM easily \citep{sene13}. First identification of methyl acetate in the ISM was reported very 
recently \citep{terc13} using IRAM $30 \ \mathrm{m}$ radio telescope towards Orion constellation. They 
reported the discovery of methyl acetate through the detection of large 
number of rotational lines from each one of the spin states of methyl acetate. 

Recently, Sivaraman et al. (2013, 2015) carried out experiments to find out the spectral behaviour 
of methyl acetate and methyl propionate 
in Vacuum Ultraviolet (VUV) and Infrared (IR) at various astrophysically
relevant temperatures (from $10 \ \mathrm{K}$ to sublimation of ice) in ultrahigh vacuum chamber.
In between $110 \ \mathrm{K}$ to $120\ \mathrm{K}$, they found an irreversible phase change 
(from amorphous to crystalline form of methyl acetate ice). Moreover, Sivaraman et al. (2014) carried out 
electron irradiation experiment on methyl acetate at $85 \ \mathrm{K}$ to find out its dissociated 
products. According to their study, $\mathrm{CO_2}$, $\mathrm{CO}$ were found to be the major 
products along with their by-products, such as, ethane ($\mathrm{C_2H_6}$) and dimethyl ether ($\mathrm{CH_3OCH_3}$). 
In addition, they found some $\mathrm{CH_3OH}$ as a by-product, 
which indicates the possibility of conversion from acetate to alcohol. 
Garrod et al. (2008) showed formation of methyl acetate in icy mantles using the gas 
grain warm-up chemical model. However, to our knowledge, no work has been reported till date 
to model deuterated forms of methyl acetate around the ISM. 
Here, in this paper, we are proposing to observe singly deuterated methyl acetate around the
sources, where methyl acetate already been observed.

Methyl acetate is an isomer of $\mathrm{C_3H_6O_2}$. It is considered to be the most 
abundant non-cyclic isomer of $\mathrm{C_3H_6O_2}$. Among other isomers of $\mathrm{C_3H_6O_2}$, 
hydroxyacetone ($\mathrm{CH_3COCH_2OH}$) and ethyl formate ($\mathrm{C_2H_5OCHO}$) are recently 
been observed in the ISM. Propionic acid ($\mathrm{CH_3CH_2COOH}$) is also an isomer of 
$\mathrm{C_3H_6O_2}$ and it is a plausible species in regions where acetic acid is found \citep{blag03}. 
Methoxyacetaldehyde ($\mathrm{CH_3OCH_2CHO}$) is another isomer which could also be observed in the ISM.
Tercero et al. (2013) predicted an upper limit on column densities of propionic acid
and methoxyacetaldehyde to be $1.6 \times 10^{14}$ $\mathrm{cm^{-2}}$ 
and $2 \times 10^{14}$ $\mathrm{cm^{-2}}$ respectively.
A combined laboratory and astronomical investigation on the hydroxyacetone was 
conducted by Apponi et al. (2006). They derived an upper limit of $5 \times 10^{12} \ \mathrm{cm^{-2}}$ 
to the column density of hydroxyacetone in Sgr B2(N) but they pointed out that it still requires further
spectroscopic investigations before a decisive confirmation. 
Belloche et al. (2009) detected ethyl formate towards Sgr B2(N). 
They estimated column density of ethyl formate to be 
$5.4 \times 10^{16}$ $\mathrm{cm^{-2}}$. Later, Tercero et al. (2013) detected both the conformers 
(gauche and trans) of ethyl formate in Orion. So among various isomers of $\mathrm{C_3H_6O_2}$, 
further laboratory and spectral survey are required for the confirmed detection of propionic acid and
methoxyacetaldehyde. Ethyl formate and hydroxyacetone, detections of both being
confirmed in the ISM and are included in the JPL catalog 
(http://spec.jpl.nasa.gov/ftp/pub/catalog/catdir.html). 

Plan of this paper is as follows. In Section 2, chemical modeling and results are discussed. In 
Section 3, spectroscopical modeling are discussed and finally, in Section 4, 
we draw our conclusion.

\section{Chemical modeling}

We use our updated large gas-grain chemical network for the purpose of chemical modeling. 
We assume that gas and grains are coupled through accretion and thermal/non-thermal/cosmic ray evaporation 
processes. Details of these processes are already presented in Das et al. (2011, 2013a, 2014). 

\subsection{Gas phase reaction network}

Here, we consider a large gas-grain chemical network to study the chemical evolution of
methyl acetate along with its two singly deuterated isotopomers. We consider
$6277$ gas phase reactions between $628$ gas phase species. This gas phase chemical network 
principally adopted from UMIST 2006 \citep{wood07} database.
Since this database does not consider the 
ortho and para spin modifications of various $\mathrm{H}$ bearing species and also 
our quantum chemical results are nuclear spin independent, we are not considering any 
nuclear spin states in our model.
Some deuterated reactions are included by following Roberts \& Millar (2000);
Albertsson et al. (2013); Das et al. (2014). 
Moreover, we include some reactions which would eventually leads to
form two singly deuterated isotopomers and some isomers of methyl acetate. 
This additional reactions
are considered by following some earlier studies and some educated guesses. 
Gas phase reaction network for the formation/destruction of methyl acetate 
along with its deuterated isotopomers are given in Table 1.
Rate coefficients for all the reactions are shown in
Table 1. Here, we either use existing empirical relations
for the gas phase rate coefficients or assume same
rate coefficients for the reactions of comparable varieties. Rate
coefficients of some reactions are calculated indigenously from our quantum chemical 
calculations.
It is obvious that due to the zero point correction energy rate coefficients between the hydrogenated
and deuterated species differ. But due to unavailability of suitable network, we assume the same reaction rates as some 
of the hydrogenated reactions. 

Gas phase methyl acetate mainly forms in the ISM by the reaction (reaction number $R1$) 
between methoxy radical ($\mathrm{CH_3O}$) and acetyl radical ($\mathrm{CH_3CO}$). 
Electronic structure calculations are carried out by using GAUSSIAN 09W program  \citep{fore96} to compute the
rate coefficient of this reaction. DFT calculations are capable of reproducing experimental 
information with high reliableness \citep{glas07}. Here computations are performed by using 
Becke three-parameter Exchange 
and Lee, Yang and Parr correlation (B3LYP) functional \citep{beck93} with the 6-311++G(d,p) 
basis set available in the Gaussian 09W program. It is noticed that the difference between 
the sum of electronic and thermal enthalpies of products and reactants of reaction $R1$ is 
$-349.657$ $\mathrm{kJ/mol}$. Other radical-radical reactions (reaction number $R2-R4$) 
of methyl acetate network are also highly exothermic in nature, having, 
$\Delta H  =  -318.925, \ -318.917$ and $-291.832$ $\mathrm{kJ/mol}$ respectively. 
Thus, $\Delta H< 0$ for these reactions, which indicates that these reactions are highly 
exothermic in nature . Now for the highly exothermic reactions, it is expected to 
have no barriers even around 
the low temperatures \citep{smit06,smit11} and it does not follow temperature dependent 
Arrhenius expression. There are a number of statistical treatments available for the calculation 
of rate coefficients of these types of reactions \citep{herb87,bate83,wake10,geor05,bett95,bass81}. 
Here, we use the following semi empirical relationship developed by Bates (1983), which
had been used in some earlier studies \citep{gupt11,maju12}:
\begin{equation}
K= 1 \times 10^{-21} A_r (6E_0 + N -2)^(3N-7)/(3N-7)! \ cm^3 s^{-1} \,
\end{equation}
where, $E_0$ is the magnitude of association energy in $eV$, $A_r$ ($A_r=100$ unless better informations 
are available) is the transition probability (in $s^{-1}$) of stabilizing the transition and $N$ 
is the number of nuclei in the complex ($N=11$ for reaction no. $R1-R4$). 
The upper limit set by following equation is adopted only 
when the calculated value from eqn. 1 exceeds the limit set by the following equation: 
\begin{equation}
K= 7.41 \times 10^{-10} {\alpha_p}^{1/2} (10/\mu)^{1/2} \ cm^3 s^{-1} \ ,
\end{equation}
where, $\alpha_p$ is the polarizability (we use isotropic polarizability of the largest radical 
having an evenly distributed charge) in $\AA^3$ and $\mu$ is the reduced mass of the reactants 
on $12_C$ amu scale as suggested by Bates (1983). In case of reaction no. (R1-R4), $\mathrm{CH_3CO}$ 
is the largest radical having $\alpha = 3.917 \ \mathrm{\AA^3}$ \citep{lide01}.

Calculations of rate coefficients by eqn. 1 implies $1.19 \times 10^{-7}, \ 2.25 \times 10^{-8}, \
2.24 \times 10^{-8}$  and $4.70 \times 10^{-9}$  cm$^3$ s$^{-1}$ respectively for reactions 
$R1-R4$ and calculations by  eqn. 2 sets the upper limit of 
$8.44 \times 10^{-10}, \ 8.37 \times 10^{-10}, \ 8.37 \times 10^{-10}$ and  $8.44 \times 10^{-10}$ 
cm$^3$ s$^{-1}$ respectively for reactions $R1-R4$. 
Clearly the rate coefficients calculated by eqn. 1 exceeds the limit set by eqn. 2. 
Thus, we are adopting only the upper limits for these reactions.

Methyl acetate could be destroyed by various types of reactions (mainly by ion-molecular reactions, 
and interaction with cosmic rays or with interstellar photons). Since gas-phase 
chemistry of cold regions are mainly dominated by exothermic ion-molecular reactions, 
ion-molecular pathways could serve as the dominant destruction pathways for any gas 
phase neutral molecules. We consider some ion molecular reactions (given in Table 1)
for the destruction of $\mathrm{CH_3COOCH_3}$ (only some dominant ions are considered). 
All the reactions along with their rate coefficients are shown in Table 1. 
Rate constants for ion-molecular reactions can be 
predicted by using capture theories \citep{herb06}. 
But $\mathrm{CH_3COOCH_3}$ is a polar neutral species 
with a dipole moment $1.72 \ \mathrm{Debye}$. Due to the interaction between the 
charge and rotating permanent   
dipole moment, a complex situation might arise in this case. Thus, here,
we use the relation developed by Su \& Chesnavich (1982) to predict the rate coefficients ($k_{cap}$).
Woon \& Herbst (2009) wrote Su-Chesnavich formula in two different ways by using
a parameter $x= \mu_D/\sqrt(2 \alpha_p T)$, where $k$ is Boltzman constant
\& $T$ is the temperature. 
From the quantum chemical calculation, we are having polarizability of methyl acetate is
$6.93$ $\AA$. For the other isomers and deuterated forms of methyl acetate, we are considering
similar $\alpha_p$ for our calculations.

The ion-dipole ($k_{cap}$) rates can be in following form;
\begin{equation}
                    k_{cap}= (0.4767 x + 0.6200) k_L,
\end{equation}
and
\begin{equation}
               k_{cap}= [(x+0.5090)^2/10.526 + 0.9754] k_L.
\end{equation}
When $x \ge 2$, Eqn. 2 is used and when $x<2$, Eqn. 3 is used. It clear that at $x=0$, above relations
reduce to the Langevin expression. In terms of temperature, above expression could be written
alternatively into the following form. For example, when $x \geq 2$,
\begin{equation}
                            k_{cap}= c_1 + c_2 T^{-1/2},
\end{equation}
where,  $c_1= 0.62 k_L$ and $c_2= (0.4767 \mu_D/\sqrt(2 \alpha K) k_L$.

For the cosmic ray induced photo-reactions, we use following rate coefficient as
used in Woodall et al. (2007),
\begin{equation}
k_{CR} = \alpha (T /300)^{\beta} \gamma /(1 - \omega),
\end{equation}
where, $\alpha$ is the cosmic ray ionization rate, $\gamma$ is the probability 
per cosmic ray ionization that the appropriate photo reaction takes place and 
$\omega$ is the dust-grain albedo in the far-ultraviolet. We use $\alpha = 1.30 \times 10^{-17}$
and $\omega = 0.6$ by following all other cosmic ray induced photo reactions. 
$\gamma$ is highly sensitive to the reactants. 
Since no guess for this parameter was available, we choose this parameter
by following cosmic ray induced photo reactions of similar types of species.
For the cosmic ray induced photo reactions of $\mathrm{HCOOCH_3}$, $\gamma=1000$
was used in Woodall et al. (2007). We use a similar choice for $\gamma$ parameter while
calculating the cosmic ray induced rate coefficients   
for $\mathrm{CH_3COOCH_3}$ and its related species.

For the interstellar photo reactions the rate is derived by using the following relation
used in Woodall et al. (2007),
\begin{equation}
k_{PHOTON} =  \alpha exp(-\gamma A_V)  s^{-1},
\end{equation}
where, $\alpha$ is the rate in the unshielded interstellar ultraviolet radiation field and 
$\gamma$ is used to control the increased extinction of dust at the ultraviolet wavelength.
Here, we use $\alpha=1.0 \times 10^{-9}$ and $\gamma=1.0$.
All the reaction types mentioned above are given in Table 1 for $T=10$ K and $A_V=10$.

\subsection{Surface reaction network}

Chemical enrichment of interstellar grain mantle solely depends on the binding energies of surface 
species (Das \& Chakrabarti, 2011). Mobility of lighter species such as $\mathrm{H, \ D, \ N}$ and $\mathrm{O}$ mainly 
dictates chemical composition of interstellar grain mantle. Composition of grain mantle
which depends on mobility of $\mathrm{H}$ and $\mathrm{O}$ atoms are already discussed in Das et al. (2008a);
Das et al. (2010) and Das \& Chakrabarti (2011). We assume that the gas phase 
species are physisorbed onto dust grains ($\sim 0.1 \ \mathrm{\mu m}$) having a grain number density of 
$1.33 \times 10^{-12} n_H$, where $n_H$ is the concentration of $\mathrm{H}$ nuclei in all forms. 
Due to the unavailability of the binding energies of deuterated species, we assume that it is 
the same as their hydrogenated counterparts. Binding energies of all surface species are mainly adopted from 
Hasegawa \& Herbst (1993); Allen \& Robindon (1977). For some species, we use binding energies from 
the latest modeling by Garrod et al. (2008) (use $ED_{\mathrm{CH_3}}=588\ \mathrm{K}, 
ED_{\mathrm{HCO/DCO}}=800 \ \mathrm{K}, ED_{\mathrm{CH_3O/CH_2DO}}=1250 \ \mathrm{K}, 
ED_{\mathrm{CH_2OH/CH_2OD}}=2254 \ \mathrm{K}$ from Garrod et al. 2008). 
Since the binding energy of $\mathrm{CH_3CO}$ was unavailable. By following the binding energy of
$\mathrm{CH_2CO}$ in Allen \& Robinson (1977), we consider the binding energy of 
$\mathrm{CH_3CO}$ to be $2520$ K.

To construct the surface reaction network, we primarily follow Hasegawa et al. (1992);
Cuppen \& Herbst (2007) and Das et al. (2010, 2011, 2014).
For the deuterium fractionation reactions on a grain surface, we primarily follow 
Caselli (2002); Cazau et al. (2010); Garrod et al. (2008)  and Das et al. (2014). 
Formation and destruction pathways of methyl acetate along with its singly deuterated isotopomers 
and its two isomers are given in Table 2. We also include 
photo dissociations of surface species by following Garrod et al. (2008). 
Photo dissociation of methyl acetate and its deuterated species are included by following
Sivaraman et al. (2014). Various evaporation processes are adopted, namely: (a) thermal evaporation,
(b) cosmic ray induced evaporation and (c) non-thermal desorption.
Due to the lower adsorption energies of the lighter species ($\mathrm{H, \ D, \ O}$ etc.), 
thermal evaporation is very efficient. Heavier surface species mostly remain on the grain surface after their
formation or accretion from the gas phase. During the late stage of the evolution process, cosmic ray induced 
evaporation serves as an efficient means to transfer surface species into the gas phase.
Another important evaporation mechanism is the non-thermal desorption process. 
Due to the energy release during some reactions, adsorbed species could be desorbed
just after their formation. Garrod et al. (2007) estimated that a fraction `$f$' of the product 
species in qualifying reactions could desorb immediately and the rest $(1-f)$ fraction remains 
as a surface bound product. Here, we apply this mechanism to all surface reactions which result 
in a single product. Fraction `f' is calculated by $f=\frac{aP}{(1+aP)}$, where, $a$ is 
the ratio between surface molecule bond frequency to frequency at which energy is lost to
the grain surface. Here, we choose $a=0.05$ for all our surface reactions. All these evaporation processes are
discussed in detail in our earlier models \citep{maju14a,maju14b,das15}.

\begin{table*}
\centering{
\scriptsize
\caption{Reaction network of various forms of methyl acetate and its two isomers in gas phase}
\begin{tabular}{|l|c|c|}
\hline
Reaction &Reaction Type& Rate coefficients \\
&& at $T=10$K, $A_V=10$ \\
\hline\hline
$\mathrm{CH_3O + CH_3CO\rightarrow CH_3COOCH_3}$ \ (R1) & Radical-radical& $8.44 \times 10^{-10}$ $\mathrm{cm^3s^{-1}}$\\
$\mathrm{CH_2DO+ CH_3CO\rightarrow CH_3COOCH_2D}$ \ (R2) & Radical-radical& $8.37 \times 10^{-10}$ $\mathrm{cm^3s^{-1}}$\\
$\mathrm{CH_2DO+  CH_3CO\rightarrow CH_2DCOOCH_3}$ \ (R3) & Radical-radical& $8.37 \times 10^{-10}$ $\mathrm{cm^3s^{-1}}$\\
$\mathrm{CH_2OH+  CH_3CO\rightarrow CH_3COCH_2OH}$ \ (R4)& Radical-radical&  $8.44 \times 10^{-10}$ $\mathrm{cm^3s^{-1}}$\\
$\mathrm{C^++  CH_3COOCH_3\rightarrow C_3H_6O_2^+ + C}$ \ (I1)&Ion-molecular&  $1.25 \times 10^{-08}$ $\mathrm{cm^3s^{-1}}$\\
$\mathrm{H^++  CH_3COOCH_3\rightarrow C_3H_6O_2^+ + H}$ \ (I2) &Ion-molecular& $4.05 \times 10^{-08}$ $\mathrm{cm^3s^{-1}}$\\
$\mathrm{H_3^++  CH_3COOCH_3\rightarrow C_3H_7O_2^+ + H_2}$ \ (I3) &Ion-molecular& $2.37 \times 10^{-08}$ $\mathrm{cm^3s^{-1}}$\\
$\mathrm{H_3O^++  CH_3COOCH_3\rightarrow C_3H_7O_2^+ + H_2O}$ \ (I4) &Ion-molecular& $1.03 \times 10^{-08}$ $\mathrm{cm^3s^{-1}}$\\
$\mathrm{HCO^++  CH_3COOCH_3\rightarrow C_3H_7O_2^+ + CO}$ \ (I5) &Ion-molecular& $8.81 \times 10^{-09}$ $\mathrm{cm^3s^{-1}}$\\
$\mathrm{He^++  CH_3COOCH_3\rightarrow C_2H_3O_2^+ + CH_3 + He}$ \ (I6) &Ion-molecular& $2.88 \times 10^{-08}$ $\mathrm{cm^3s^{-1}}$\\
$\mathrm{He^++  CH_3COOCH_2D\rightarrow C_2H_2DO_2^+ + CH_3 + He}$ \ (I7) &Ion-molecular& $2.88 \times 10^{-08}$ $\mathrm{cm^3s^{-1}}$\\
$\mathrm{C^++  CH_3COOCH_2D\rightarrow C_3H_5DO_2^+ + C}$ \ (I8)&Ion-molecular&  $1.25 \times 10^{-08}$ $\mathrm{cm^3s^{-1}}$\\
$\mathrm{H^++  CH_3COOCH_2D\rightarrow C_3H_5DO_2^+ + H}$ \ (I9) &Ion-molecular& $4.05 \times 10^{-08}$ $\mathrm{cm^3s^{-1}}$\\
$\mathrm{H_3^++  CH_3COOCH_2D\rightarrow C_3H_6DO_2^+ + H_2}$ \ (I10) &Ion-molecular& $2.37 \times 10^{-08}$ $\mathrm{cm^3s^{-1}}$\\
$\mathrm{H_3O^++  CH_3COOCH_2D\rightarrow C_3H_6DO_2^+ + H_2O}$ \ (I11)&Ion-molecular&  $1.03 \times 10^{-08}$ $\mathrm{cm^3s^{-1}}$\\
$\mathrm{HCO^++  CH_3COOCH_2D\rightarrow C_3H_6DO_2^+ + CO}$ \ (I12) &Ion-molecular& $8.81 \times 10^{-09}$ $\mathrm{cm^3s^{-1}}$\\
$\mathrm{He^++  CH_3COOCH_2D\rightarrow C_2H_2DO_2^+ + CH_2D + He}$ \ (I13) &Ion-molecular& $2.88 \times 10^{-08}$ $\mathrm{cm^3s^{-1}}$\\
$\mathrm{C^++  CH_2DCOOCH_3\rightarrow C_3H_5DO_2^+ + C}$ \ (I14) &Ion-molecular& $1.24 \times 10^{-08}$ $\mathrm{cm^3s^{-1}}$\\
$\mathrm{H^++  CH_2DCOOCH_3\rightarrow C_3H_5DO_2^+ + H}$ \ (I15) &Ion-molecular& $4.05 \times 10^{-08}$ $\mathrm{cm^3s^{-1}}$\\
$\mathrm{H_3^++  CH_2DCOOCH_3\rightarrow C_3H_6DO_2^+ + H_2}$ \ (I16) &Ion-molecular& $2.37 \times 10^{-08}$ $\mathrm{cm^3s^{-1}}$\\
$\mathrm{H_3O^++  CH_2DCOOCH_3\rightarrow C_3H_6DO_2^+ + H_2O}$ \ (I17) &Ion-molecular& $1.03 \times 10^{-08}$ $\mathrm{cm^3s^{-1}}$\\
$\mathrm{HCO^++  CH_2DCOOCH_3\rightarrow C_3H_6DO_2^+ + CO}$ \ (I18) &Ion-molecular& $8.81 \times 10^{-09}$ $\mathrm{cm^3s^{-1}}$\\
$\mathrm{H^++  C_2H_6\rightarrow C_2H_5^+ + H_2}$ \ (I19) &Ion-molecular& $1.65 \times 10^{-09}$ $\mathrm{cm^3s^{-1}}$\\
$\mathrm{He^++  C_2H_6\rightarrow C_2H_5^+ + He + H}$ \ (I20) &Ion-molecular& $5.00 \times 10^{-09}$ $\mathrm{cm^3s^{-1}}$\\
$\mathrm{H_3^++  CH_3OCH_2D\rightarrow CH_3OCH_3D^+ + H_2}$ \ (I21) &Ion-molecular& $2.00 \times 10^{-09}$ $\mathrm{cm^3s^{-1}}$\\
$\mathrm{He^++  CH_3OCH_2D\rightarrow H_2CO + CH_2D^+ + He + H}$ \ (I22) &Ion-molecular& $2.00 \times 10^{-09}$ $\mathrm{cm^3s^{-1}}$\\
$\mathrm{CH_3^++  CH_3OCH_2D\rightarrow CH_3CHOH^+ + CH_3D}$ \ (I23) &Ion-molecular& $3.50 \times 10^{-10}$ $\mathrm{cm^3s^{-1}}$\\
$\mathrm{H_3O^++  CH_3OCH_2D\rightarrow CH_3OCH_3D^+  +HDO}$ \ (I24) &Ion-molecular& $2.70 \times 10^{-09}$ $\mathrm{cm^3s^{-1}}$\\
$\mathrm{C_2H_5OH_2^++  H_2CO\rightarrow HCOOC_2{H_6}^+ + H_2}$ \ (I25) &Ion-molecular& $1.24 \times 10^{-08}$ $\mathrm{cm^3s^{-1}}$\\
$\mathrm{C^++  C_2H_5OCHO\rightarrow C_3H_6O_2^+ + C}$ \ (I26) &Ion-molecular& $1.25 \times 10^{-08}$ $\mathrm{cm^3s^{-1}}$\\
$\mathrm{C^++  CH_3COCH_2OH\rightarrow C_3H_6O_2^+ + C}$ \ (I27) &Ion-molecular& $1.25 \times 10^{-08}$ $\mathrm{cm^3s^{-1}}$\\
$\mathrm{H^++  C_2H_5OCHO\rightarrow C_3H_6O_2^+ + H}$ \ (I28) &Ion-molecular& $4.05 \times 10^{-08}$ $\mathrm{cm^3s^{-1}}$\\
$\mathrm{H^++  CH_3COCH_2OH\rightarrow C_3H_6O_2^+ + H}$ \ (I29) &Ion-molecular& $4.05 \times 10^{-08}$ $\mathrm{cm^3s^{-1}}$\\
$\mathrm{H_3^++  C_2H_5OCHO\rightarrow C_3H_7O_2^+ + H_2}$ \ (I30) &Ion-molecular& $2.37 \times 10^{-08}$ $\mathrm{cm^3s^{-1}}$\\
$\mathrm{H_3^++  CH_3COCH_2OH\rightarrow C_3H_7O_2^+ + H_2}$ \ (I31) &Ion-molecular& $2.37 \times 10^{-08}$ $\mathrm{cm^3s^{-1}}$\\
$\mathrm{H_3O^++  C_2H_5OCHO\rightarrow C_3H_7O_2^+ + H_2O}$ \ (I32) &Ion-molecular& $1.03 \times 10^{-08}$ $\mathrm{cm^3s^{-1}}$\\
$\mathrm{H_3O^++  CH_3COCH_2OH\rightarrow C_3H_7O_2^+ + H_2O}$ \ (I33) &Ion-molecular& $1.03 \times 10^{-08}$ $\mathrm{cm^3s^{-1}}$\\
$\mathrm{HCO^++  C_2H_5OCHO\rightarrow C_3H_7O_2^+ + CO}$ \ (I34) &Ion-molecular& $8.81 \times 10^{-09}$ $\mathrm{cm^3s^{-1}}$\\
$\mathrm{HCO^++  CH_3COCH_2OH\rightarrow C_3H_7O_2^+ + CO}$ \ (I35) &Ion-molecular& $8.81 \times 10^{-09}$ $\mathrm{cm^3s^{-1}}$\\
$\mathrm{He^++  C_2H_5OCHO\rightarrow C_2H_3O_2^+ + CH_3+ He}$ \ (I36) &Ion-molecular& $2.88 \times 10^{-08}$ $\mathrm{cm^3s^{-1}}$\\
$\mathrm{He^++  CH_3COCH_2OH\rightarrow C_2H_3O_2^+ + CH_3 He}$ \ (I37) &Ion-molecular& $2.88 \times 10^{-08}$ $\mathrm{cm^3s^{-1}}$\\
$\mathrm{C^++  C_2H_4DOCHO\rightarrow C_3H_5DO_2^+ + C}$ \ (I38) &Ion-molecular& $1.25 \times 10^{-08}$ $\mathrm{cm^3s^{-1}}$\\
$\mathrm{H^++  C_2H_4DOCHO\rightarrow C_3H_5DO_2^+ + H}$ \ (I39) &Ion-molecular& $4.05 \times 10^{-08}$ $\mathrm{cm^3s^{-1}}$\\
$\mathrm{H_3^++  C_2H_4DOCHO\rightarrow C_3H_6DO_2^+ + H_2}$ \ (I40) &Ion-molecular& $2.37 \times 10^{-08}$ $\mathrm{cm^3s^{-1}}$\\
$\mathrm{H_3O^++  C_2H_4DOCHO\rightarrow C_3H_6DO_2^+ + H_2O}$ \ (I42) &Ion-molecular& $1.03 \times 10^{-08}$ $\mathrm{cm^3s^{-1}}$\\
$\mathrm{HCO^++  C_2H_4DOCHO\rightarrow C_3H_6DO_2^+ + CO}$ \ (I43) &Ion-molecular& $8.81 \times 10^{-09}$ $\mathrm{cm^3s^{-1}}$\\
$\mathrm{He^++  C_2H_4DOCHO\rightarrow C_2H_3O_2^+ + CH_2D+ He}$ \ (I44) &Ion-molecular& $2.88 \times 10^{-08}$ $\mathrm{cm^3s^{-1}}$\\
$\mathrm{H_2D^++  CH_3COOCH_3\rightarrow C_3H_7O_2^+ + HD}$ \ (I45) &Ion-molecular& $2.37 \times 10^{-08}$ $\mathrm{cm^3s^{-1}}$\\
$\mathrm{H_2D^++  CH_3COOCH_3\rightarrow C_3H_6DO_2^+ + H_2}$ \ (I46) &Ion-molecular& $2.37 \times 10^{-08}$ $\mathrm{cm^3s^{-1}}$\\
$\mathrm{H_2D^++  CH_3COOCH_2D\rightarrow C_3H_6DO_2^+ + HD}$ \ (I47) &Ion-molecular& $2.37 \times 10^{-08}$ $\mathrm{cm^3s^{-1}}$\\
$\mathrm{H_2D^++  CH_2DCOOCH_3\rightarrow C_3H_6DO_2^+ + HD}$ \ (I48) &Ion-molecular& $2.37 \times 10^{-08}$ $\mathrm{cm^3s^{-1}}$\\
$\mathrm{H^++  C_2H_6\rightarrow C_2H_4^+ + H_2 + H}$ \ (C1) &Charge Exchange& $3.06 \times 10^{-09}$ $\mathrm{cm^3s^{-1}}$\\
$\mathrm{He^++  C_2H_6\rightarrow C_2H_6^+ + He}$ \ (C2) &Charge Exchange& $5.00 \times 10^{-10}$ $\mathrm{cm^3s^{-1}}$\\
$\mathrm{C^++  C_2H_6\rightarrow C_2H_6^+ + C}$ \ (C3) &Charge Exchange& $5.00 \times 10^{-10}$ $\mathrm{cm^3s^{-1}}$\\
$\mathrm{H^++  CH_3OCH_2D\rightarrow CH_3OCH_2D^+ + H}$ \ (C4) &Charge Exchange& $2.00 \times 10^{-09}$ $\mathrm{cm^3s^{-1}}$\\
$\mathrm{C^++  CH_3OCH_2D\rightarrow CH_3OCH_2D^+ + C}$ \ (C5) &Charge Exchange& $2.00 \times 10^{-09}$ $\mathrm{cm^3s^{-1}}$\\
$\mathrm{C_2H_3O_2^++  e-\rightarrow CO_2 + CH_3}$ \ (D1) &Dissociative Recombination& $1.60 \times 10^{-06}$ $\mathrm{cm^3s^{-1}}$\\
$\mathrm{C_2H_2DO_2^++  e-\rightarrow CO_2 + CH_2D}$ \ (D2) &Dissociative Recombination& $1.60 \times 10^{-06}$ $\mathrm{cm^3s^{-1}}$\\
$\mathrm{C_3H_6O_2^++  e-\rightarrow HCOOCH_3 + CH_2}$ \ (D3) &Dissociative Recombination& $8.21 \times 10^{-07}$ $\mathrm{cm^3s^{-1}}$\\
$\mathrm{C_3H_5DO_2^++  e-\rightarrow HCOOCH_3 + CHD}$ \ (D4) &Dissociative Recombination& $8.21 \times 10^{-07}$ $\mathrm{cm^3s^{-1}}$\\
$\mathrm{C_3H_7O_2^++  e-\rightarrow HCOOCH_3 + CH_3}$ \ (D5) &Dissociative Recombination& $8.21 \times 10^{-07}$ $\mathrm{cm^3s^{-1}}$\\
$\mathrm{C_3H_6DO_2^++  e-\rightarrow HCOOCH_3 + CH_2D}$ \ (D6) &Dissociative Recombination& $8.21 \times 10^{-07}$ $\mathrm{cm^3s^{-1}}$\\
$\mathrm{C_2H_6^++  e-\rightarrow C_2H_2 + H_2 + H_2}$ \ (D7) &Dissociative Recombination& $1.19\times 10^{-06}$ $\mathrm{cm^3s^{-1}}$\\
$\mathrm{C_2H_5D^++  e-\rightarrow C_2H_2 + HD + H_2}$ \ (D8) &Dissociative Recombination& $1.19 \times 10^{-06}$ $\mathrm{cm^3s^{-1}}$\\
$\mathrm{C_2H_6^++  e-\rightarrow C_2H_4 + H_2}$ \ (D9) &Dissociative Recombination& $4.99 \times 10^{-07}$ $\mathrm{cm^3s^{-1}}$\\
$\mathrm{C_2H_5D^++  e-\rightarrow C_2H_4 + HD}$ \ (D10) &Dissociative Recombination& $4.99 \times 10^{-07}$ $\mathrm{cm^3s^{-1}}$\\
$\mathrm{C_2H_6^++  e-\rightarrow C_2H_4 + H + H}$ \ (D11) &Dissociative Recombination& $1.11 \times 10^{-06}$ $\mathrm{cm^3s^{-1}}$\\
$\mathrm{C_2H_5D^++  e-\rightarrow C_2H_4 + D + H}$ \ (D12) &Dissociative Recombination& $1.11 \times 10^{-06}$ $\mathrm{cm^3s^{-1}}$\\
$\mathrm{C_2H_6^++  e-\rightarrow CH_3 + CH_3}$ \ (D13) &Dissociative Recombination& $7.04 \times 10^{-07}$ $\mathrm{cm^3s^{-1}}$\\
$\mathrm{C_2H_5D^++  e-\rightarrow CH_2D + CH_3}$ \ (D14) &Dissociative Recombination& $7.04 \times 10^{-07}$ $\mathrm{cm^3s^{-1}}$\\
$\mathrm{C_2H_6^++  e-\rightarrow C_2H_3 + H + H + H}$ \ (D15) &Dissociative Recombination& $5.29 \times 10^{-07}$ $\mathrm{cm^3s^{-1}}$\\
\hline
\end{tabular}}
\end{table*}

\clearpage
\clearpage
\hskip 1cm
{\scriptsize
\centering
\begin{tabular}{|l|c|c|}
\hline
$\mathrm{C_2H_5D^++  e-\rightarrow C_2H_3 + D + H + H}$ \ (D16) &Dissociative Recombination& $5.29 \times 10^{-07}$ $\mathrm{cm^3s^{-1}}$\\
$\mathrm{CH_3OCH_3D^++  e-\rightarrow CH_3OCH_2D + H}$ \ (D17) &Dissociative Recombination& $8.22 \times 10^{-07}$ $\mathrm{cm^3s^{-1}}$\\
$\mathrm{CH_3OCH_2D^++  e-\rightarrow O + CH_3 + CH_2D}$ \ (D18) &Dissociative Recombination& $8.22 \times 10^{-07}$ $\mathrm{cm^3s^{-1}}$\\
$\mathrm{CH_3OCH_2D^++  e-\rightarrow CH_3OD + CH_2}$ \ (D19) &Dissociative Recombination& $8.22 \times 10^{-07}$ $\mathrm{cm^3s^{-1}}$\\
$\mathrm{HCOOC_2{H_6}^++  e-\rightarrow C_2H_5OCHO + H}$ \ (D20) &Dissociative Recombination& $1.56 \times 10^{-06}$ $\mathrm{cm^3s^{-1}}$\\
$\mathrm{C_3H_7{O_2}^++  e-\rightarrow C_2H_5OCHO + H}$ \ (D21) &Dissociative Recombination& $1.56 \times 10^{-06}$ $\mathrm{cm^3s^{-1}}$\\
$\mathrm{C_3H_7{O_2}^++  e-\rightarrow CH_3COOCH_3 + H}$ \ (D22) &Dissociative Recombination& $1.56 \times 10^{-06}$ $\mathrm{cm^3s^{-1}}$\\
$\mathrm{C_3H_7{O_2}^++  e-\rightarrow CH3COCH_2OH + H}$ \ (D23) &Dissociative Recombination& $1.56 \times 10^{-06}$ $\mathrm{cm^3s^{-1}}$\\
$\mathrm{C_3H_6D{O_2}^++  e-\rightarrow C_2H_4DOCHO + H}$ \ (D24) &Dissociative Recombination& $1.56 \times 10^{-06}$ $\mathrm{cm^3s^{-1}}$\\
$\mathrm{CH_3COOCH_3+  CRPHOT\rightarrow C_3H_6O_2^+ + e-}$ \ (C1) &Cosmic ray induced photoreaction& $3.25 \times 10^{-14}$ $\mathrm{s^{-1}}$\\
$\mathrm{C_2H_5OCHO+  CRPHOT\rightarrow C_3H_6O_2^+ + e-}$ \ (C2) &Cosmic ray induced photoreaction& $3.25 \times 10^{-14}$ $\mathrm{s^{-1}}$\\
$\mathrm{CH_3COCH_2OH + CRPHOT\rightarrow C_3H_6O_2^+ + e-}$ \ (C3) &Cosmic ray induced photoreaction& $3.25 \times 10^{-14}$ $\mathrm{s^{-1}}$\\
$\mathrm{CH_3COOCH_2D+  CRPHOT\rightarrow C_3H_5DO_2^+ + e-}$ \ (C4) &Cosmic ray induced photoreaction& $3.25 \times 10^{-14}$ $\mathrm{s^{-1}}$\\
$\mathrm{CH_2DCOOCH_3+  CRPHOT\rightarrow C_3H_5DO_2^+ + e-}$ \ (C5) &Cosmic ray induced photoreaction& $3.25 \times 10^{-14}$ $\mathrm{s^{-1}}$\\
$\mathrm{CH_3COOCH_3+  CRPHOT\rightarrow C_2H_6 + CO_2}$ \ (C6) &Cosmic ray induced photoreaction& $1.63 \times 10^{-14}$ $\mathrm{s^{-1}}$\\
$\mathrm{C_2H_5OCHO+  CRPHOT\rightarrow C_2H_6 + CO_2}$ \ (C7) &Cosmic ray induced photoreaction& $1.63 \times 10^{-14}$ $\mathrm{s^{-1}}$\\
$\mathrm{CH_3COCH_2OH+  CRPHOT\rightarrow C_2H_6 + CO_2}$ \ (C8) &Cosmic ray induced photoreaction& $1.63 \times 10^{-14}$ $\mathrm{s^{-1}}$\\
$\mathrm{CH_3COOCH_2D+  CRPHOT\rightarrow C_2H_5D + CO_2}$ \ (C9)&Cosmic ray induced photoreaction& $1.63 \times 10^{-14}$  $\mathrm{s^{-1}}$\\
$\mathrm{CH_2DCOOCH_3+  CRPHOT\rightarrow C_2H_5D + CO_2}$ \ (C10) &Cosmic ray induced photoreaction& $1.63 \times 10^{-14}$ $\mathrm{s^{-1}}$\\
$\mathrm{C_2H_6+  CRPHOT\rightarrow C_2H_6^+ + e-}$ \ (C11) &Cosmic ray induced photoreaction& $1.26 \times 10^{-14}$ $\mathrm{s^{-1}}$\\
$\mathrm{C_2H_5D+  CRPHOT\rightarrow C_2H_5D^+ + e-}$ \ (C12) &Cosmic ray induced photoreaction& $1.26 \times 10^{-14}$ $\mathrm{s^{-1}}$\\
$\mathrm{C_2H_5D+  CRPHOT\rightarrow C_2H_4 + HD}$ \ (C13) &Cosmic ray induced photoreaction& $6.11 \times 10^{-14}$ $\mathrm{s^{-1}}$\\
$\mathrm{C_2H_4DOCHO+  CRPHOT\rightarrow C_3H_5DO_2^+ + e-}$ \ (C14) &Cosmic ray induced photoreaction& $3.25 \times 10^{-14}$ $\mathrm{s^{-1}}$\\
$\mathrm{C_2H_4DOCHO+  CRPHOT\rightarrow C_3H_5D + CO_2}$ \ (C15) &Cosmic ray induced photoreaction& $3.25 \times 10^{-14}$ $\mathrm{s^{-1}}$\\
$\mathrm{CH_3COOCH_3+  PHOTON\rightarrow C_2H_6 + CO_2}$ \ (P1) &Photoreactions& $4.54 \times 10^{-14}$ $\mathrm{s^{-1}}$\\
$\mathrm{C_2H_5OCHO+  PHOTON\rightarrow C_2H_6 + CO_2}$ \ (P2) &Photoreactions& $4.54 \times 10^{-14}$ $\mathrm{s^{-1}}$\\
$\mathrm{CH_3COOCH_2D+  PHOTON\rightarrow C_2H_5D + CO_2}$ \ (P3) &Photoreactions& $4.54 \times 10^{-14}$ $\mathrm{s^{-1}}$\\
$\mathrm{CH_2DCOOCH_3+  PHOTON\rightarrow C_2H_5D + CO_2}$ \ (P4) &Photoreactions& $4.54 \times 10^{-14}$ $\mathrm{s^{-1}}$\\
$\mathrm{CH_3COOCH_3+  PHOTON\rightarrow CH_3OCH_3 + CO}$ \ (P5) &Photoreactions& $4.54 \times 10^{-14}$ $\mathrm{s^{-1}}$\\
$\mathrm{C_2H_5OCHO+  PHOTON\rightarrow CH_3OCH_3 + CO}$ \ (P6) &Photoreactions& $4.54 \times 10^{-14}$ $\mathrm{s^{-1}}$\\
$\mathrm{CH_3COCH_2OH+  PHOTON\rightarrow CH_3OCH_3 + CO}$ \ (P7) &Photoreactions& $4.54 \times 10^{-14}$ $\mathrm{s^{-1}}$\\
$\mathrm{CH_3COOCH_2D+  PHOTON\rightarrow CH_3OCH_2D + CO}$ \ (P8) &Photoreactions& $4.54 \times 10^{-14}$ $\mathrm{s^{-1}}$\\
$\mathrm{CH_2DCOOCH_3+  PHOTON\rightarrow CH_3OCH_2D + CO}$ \ (P9) &Photoreactions& $4.54 \times 10^{-14}$ $\mathrm{s^{-1}}$\\
$\mathrm{CH_3COOCH_3+  PHOTON\rightarrow CH_3OH + H_2CCO}$ \ (P10) &Photoreactions& $4.54 \times 10^{-14}$ $\mathrm{s^{-1}}$\\
$\mathrm{C_2H_5OCHO+  PHOTON\rightarrow CH_3OH + H_2CCO}$ \ (P11) &Photoreactions& $4.54 \times 10^{-14}$ $\mathrm{s^{-1}}$\\
$\mathrm{CH_3COCH_2OH+  PHOTON\rightarrow CH_3OH + H_2CCO}$ \ (P12) &Photoreactions& $4.54 \times 10^{-14}$ $\mathrm{s^{-1}}$\\
$\mathrm{CH_2DCOOCH_3+  PHOTON\rightarrow CH_2DOH + H_2CCO}$ \ (P13) &Photoreactions& $4.54 \times 10^{-14}$ $\mathrm{s^{-1}}$\\
$\mathrm{CH_3COOCH_2D+  PHOTON\rightarrow CH_2DOH + H_2CCO}$ \ (P14) &Photoreactions& $4.54 \times 10^{-14}$ $\mathrm{s^{-1}}$\\
$\mathrm{CH_3COOCH_3+  PHOTON\rightarrow CH_3O + CH_3CO}$ \ (P15) &Photoreactions& $4.54 \times 10^{-14}$ $\mathrm{s^{-1}}$\\
$\mathrm{C_2H_5OCHO+  PHOTON\rightarrow CH_3O + CH_3CO}$ \ (P16) &Photoreactions& $4.54 \times 10^{-14}$ $\mathrm{s^{-1}}$\\
$\mathrm{CH_3COCH_2OH+  PHOTON\rightarrow CH_2OH + CH_3CO}$ \ (P17) &Photoreactions& $4.54 \times 10^{-14}$ $\mathrm{s^{-1}}$\\
$\mathrm{CH_3COCH_2OH+  PHOTON\rightarrow CH_2OH + CH_3CO}$ \ (P18) &Photoreactions& $4.54 \times 10^{-14}$ $\mathrm{s^{-1}}$\\
$\mathrm{CH_3COOCH_2D+  PHOTON\rightarrow CH_2DO + CH_3CO}$ \ (P19) &Photoreactions& $4.54 \times 10^{-14}$ $\mathrm{s^{-1}}$\\
$\mathrm{CH_2DCOOCH_3+  PHOTON\rightarrow CH_2DO + CH_3CO}$ \ (P20) &Photoreactions& $4.54 \times 10^{-14}$ $\mathrm{s^{-1}}$\\
$\mathrm{CH_3CO+  PHOTON\rightarrow CH_3 + CO}$ \ (P21) &Photoreactions& $4.54 \times 10^{-14}$ $\mathrm{s^{-1}}$\\
$\mathrm{C_2H_6+  PHOTON\rightarrow C_2H_6^+ + e-}$ \ (P22) &Photoreactions& $2.78 \times 10^{-21}$ $\mathrm{s^{-1}}$\\
$\mathrm{C_2H_6+  PHOTON\rightarrow C_2H_4 + H_2}$ \ (P23) &Photoreactions& $4.14 \times 10^{-17}$ $\mathrm{s^{-1}}$\\
$\mathrm{CH_3OCH_2D+  PHOTON\rightarrow H_2CO + CH_3D}$ \ (P24) &Photoreactions& $4.72 \times 10^{-20}$ $\mathrm{s^{-1}}$\\
$\mathrm{CH_3OCH_2D+  PHOTON\rightarrow CH_3OCH_2D^+ + e-}$ \ (P25) &Photoreactions& $4.72 \times 10^{-20}$ $\mathrm{s^{-1}}$\\
$\mathrm{C_2H_4DOCHO+  PHOTON\rightarrow C_2H_5D + CO_2}$ \ (P26) &Photoreactions& $4.54 \times 10^{-14}$ $\mathrm{s^{-1}}$\\
$\mathrm{C_2H_4DOCHO+  PHOTON\rightarrow CH_3OCH_2D + CO}$ \ (P27) &Photoreactions& $4.54 \times 10^{-14}$ $\mathrm{s^{-1}}$\\
$\mathrm{C_2H_4DOCHO+  PHOTON\rightarrow CH_2DOH + H_2CCO}$ \ (P28) &Photoreactions& $4.54 \times 10^{-14}$ $\mathrm{s^{-1}}$\\
$\mathrm{C_2H_4DOCHO+  PHOTON\rightarrow CH_2DO + CH_3CO}$ \ (P29) &Photoreactions& $4.54 \times 10^{-14}$ $\mathrm{s^{-1}}$\\
$\mathrm{C_2H_5OCHO+  PHOTON\rightarrow CH_2OH + CH_3CO}$ \ (P30) &Photoreactions& $4.54 \times 10^{-14}$ $\mathrm{s^{-1}}$\\
\hline
\end{tabular}}
\clearpage

\begin{table*}
\centering{
\scriptsize
\caption{Formation/destruction of various forms of methyl acetate and its two isomers in ice phase}
\begin{tabular}{|c|c|}
\hline
Reaction & Rate coefficients at $10 \ \mathrm{K}$($\mathrm{sec^{-1}}$)\\
\hline\hline
$\mathrm{CH_3     +  CO     \rightarrow    CH_3CO       }$& $ 7.91 \times 10^{-03}$ \\
$\mathrm{CH_3O    +CH_3CO    \rightarrow    CH_3COOCH_3 }$& $ 2.31 \times 10^{-10}$ \\
$\mathrm{CH_2DO   +  CH_3CO \rightarrow    CH_2DCOOCH_3 }$& $ 2.30 \times 10^{-10}$ \\
$\mathrm{CH_2OH  +   CH_3    \rightarrow   C_2H_5OH     }$& $7.91 \times 10^{-03}$\\
$\mathrm{H      +   C_2H_5O  \rightarrow   C_2H_5OH     }$& $4.91 \times 10^{+01}$\\
$\mathrm{HCO    +   C_2H_5O  \rightarrow   C_2H_5OCHO   }$& $1.14 \times 10^{-05}$\\
$\mathrm{H       +  CH_2OCHO \rightarrow   CH_3OCHO     } $& $4.91 \times 10^{+01}$\\
$\mathrm{CH_3     +  CH_2OCHO \rightarrow  C_2H_5OCHO   }$& $7.91 \times 10^{-03}$\\
$\mathrm{HCO     +  CH_3O   \rightarrow    CH_3OCHO     } $& $1.15 \times 10^{-05}$\\
$\mathrm{CH_2OH    + CH_3CO  \rightarrow   CH_3COCH_2OH }$& $2.12 \times 10^{-10}$\\ 
$\mathrm{C_2H_5OH +   CRPHOT \rightarrow   C_2H_5O +   H}$& $1.30 \times 10^{-17}$\\
$\mathrm{C_2H_5OCHO+  CRPHOT \rightarrow   CH_2OCHO + CH_3}$&$1.30 \times 10^{-17}$\\
$\mathrm{C_2H_5OCHO+  CRPHOT  \rightarrow  C_2H_5O   +  HCO}$&$1.30 \times 10^{-17}$\\
$\mathrm{CH_3OCHO +  CRPHOT  \rightarrow   CH_2OCHO +  H  }$&$1.30 \times 10^{-17}$\\
$\mathrm{CH_3COOCH_3+ CRPHOT \rightarrow   C_2H_6    +  CO_2}$&$1.30 \times 10^{-17}$\\
$\mathrm{C_2H_5OCHO + CRPHOT \rightarrow   C_2H_6    +  CO_2}$&$1.30 \times 10^{-17}$\\
$\mathrm{CH_3COOCH_2D+CRPHOT \rightarrow   C_2H_5D   +  CO_2}$&$1.30 \times 10^{-17}$\\
$\mathrm{CH_2DCOOCH_3+CRPHOT \rightarrow   C_2H_5D   +  CO_2}$&$1.30 \times 10^{-17}$\\
$\mathrm{CH_3COCH_2OH+CRPHOT \rightarrow   C_2H_5D   +  CO_2}$&$1.30 \times 10^{-17}$\\
\hline
\end{tabular}}
\end{table*}


\begin{table*}
\centering{
\scriptsize
\caption{Initial abundances used relative to total hydrogen nuclei.}
\begin{tabular}{|c|c|}
\hline
Species&Abundance\\
\hline\hline
$\mathrm{H_2}$ &    $5.00 \times 10^{-01}$\\
$\mathrm{He}$    &    $1.00 \times 10^{-01}$\\
$\mathrm{N}$     &    $2.14 \times 10^{-05}$\\
$\mathrm{O}$     &    $1.76 \times 10^{-04}$\\
$\mathrm{H_3}$$^+$&    $1.00 \times 10^{-11}$\\
$\mathrm{C^+}$ &    $7.30 \times 10^{-05}$\\
$\mathrm{S^+}$ &    $8.00 \times 10^{-08}$\\
$\mathrm{Si^+}$&    $8.00 \times 10^{-09}$\\
$\mathrm{Fe^+}$&    $3.00 \times 10^{-09}$\\
$\mathrm{Na^+}$&    $2.00 \times 10^{-09}$\\
$\mathrm{Mg^+}$&    $7.00 \times 10^{-09}$\\
$\mathrm{P^+}$ &    $3.00 \times 10^{-09}$\\
$\mathrm{Cl^+}$&    $4.00 \times 10^{-09}$\\
$\mathrm{e^-}$ &    $7.31 \times 10^{-05}$\\
$\mathrm{HD}$&  $ 1.6 \times 10^{-05}$\\
\hline
\end{tabular}}
\end{table*}

\subsection{Physical condition}
Chemical composition of a molecular cloud is heavily dependent on the physical properties of
the cloud. Thus it is essential to consider suitable physical condition for astrochemical modeling.
Viti et al. (2004); Garrod \& Herbst (2006) \& Garrod et al. (2008) considered an isothermal 
($T=10$ K) collapsing cloud (collapsing from  $n_H=3 \times 10^3$ cm$^{-3}$ to $n_H=10^7$ cm$^{-3}$) 
for their astrochemical modeling. Here also, we consider similar collapsing cloud model 
and consider that this collapsing process would continue up to $10^6$ years. 
In Fig. 1, we show the time evolution of densities. 

Survival of complex molecules in and around any molecular cloud is strongly related 
to the effect of interstellar radiation field and thus upon the visual extinction ($A_V$) parameter.
Depending on the number density of the cloud, we vary visual extinction parameter by following 
the relation used in Lee at al. (1996) and Das \& Chakrabarti (2011). 
Relationship between the number density of hydrogen and visual extinction is given by,
\begin{equation}
\mathrm{n_H=n_{H0}[1+(\sqrt{n_{Hmax}/n_{H0}}-1)A_V/A_{Vmax}]^2},
\end{equation}
where,  $n_{H0}$ is the cloud density at the cloud surface ($= 3 \times 10^3$ cm$^{-3}$), 
$n_{Hmax}$ is the maximum density ($= 10^7$ $cm^{-3}$) and $A_{Vmax}$ ($=150$) is the 
maximum visual extinction considered very deep inside the cloud. Variation of 
$A_V$ with respect to $n_H$ is shown in Fig. 1a. 

In order to consider more realistic situation, following the consideration of Viti et al. (2004);
Garrod \& Herbst (2006) and Garrod et al. (2008), here, we use two phase physical model to mimic  
the evolution of the interstellar condition. First phase is the collapsing phase what we just have discussed 
above and second phase is the warm up phase. Collapsing phase continues up to $10^6$ year and 
in the warm up phase, it is assumed that the collapse has just stopped and temperature started to 
rise from $10$ K and would reach $200$ K. This warm up phase will continue up to another $10^6$ years.
This is the typical time scale for the low mass star formation \citep{viti04}. Fig. 1b shows the
time evolution of temperatures during our simulation regime. So in brief, our simulation 
time scale continues for $2 \times 10^6$ years -- the collapsing phase
takes $10^6$ years and the warm up phase takes up another $10^6$ years. 

\begin{figure}
\vskip 1cm
\centering{
\vbox{
  \includegraphics[height=7cm,width=8cm]{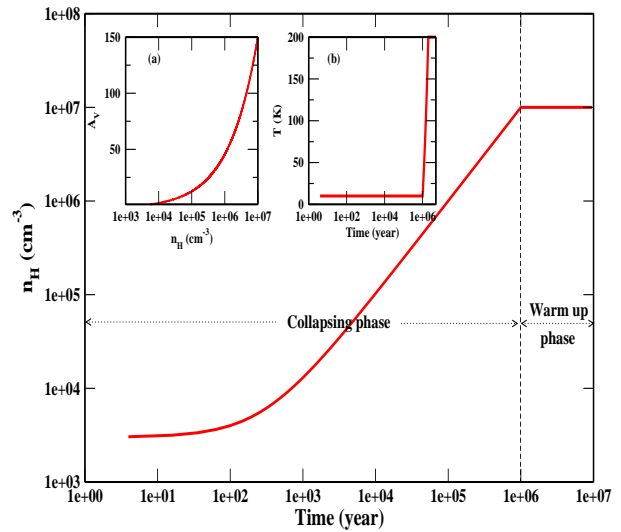}}}
\caption{\small Time evolution of densities during collapsing and warm up phase of a cloud. 
Fig. 1a (inset picture) represent the variation of visual extiction ($A_V$) with respect to the hydrogen 
number density ($n_H$) and Fig. 1b (inset picture) represent the time 
dependent temperature profile of the cloud.}
\end{figure}

\subsection{Results of chemical modeling}

Here, we consider a molecular gas to begin with. For this, we need to choose 
initial constituents relevant to a dense cloud. Our adopted initial composition
is shown in Table 3. This condition is adopted by following 
Das et al. (2014) and references therein. As in the case of hydrogen atom, it is assumed that initially all 
deuteriums are locked in the form of $\mathrm{HD}$. Initial atomic $\mathrm{D/H}$ ratio is assumed to be $0.1$.

Armed with our physical model (presented in Fig. 1) and chemical model as discussed in Section 2, 
we show time evolution of the abundances of 
$\mathrm{CH_3COOCH_3, \ CH_2DCOOCH_3/CH_3COOCH_2D}$, $\mathrm{C_2H_5COOCH_3}$ 
and $\mathrm{CH_3COCH_2OH}$ in gas phase and ice phase in Fig. 2. 
Solid curves in Fig. 2 are used to represent the gas phase species and dashed curves are used to 
represent the ice phase species. 

We are considering two phase model. During the collapsing phase, cloud collapses
isothermally ($T=10$ K). Due to low temperature in the collapsing phase, 
several species are formed on the ice phase.
It is clear from Fig. 2 that methyl acetate and its related species are mainly formed
on the ice phase during this collapsing period. In the gas phase, these molecules are produced
sufficiently but due to the depletion of gas phase species, their net production is not high.
At low temperatures, depleted species are unable to overcome the 
energy barriers on the grain surfaces and thus mostly remain trapped on the interstellar ices.
Under this circumstances, cosmic ray induced desorption and non-thermal desorption mechanisms 
discussed in the Section 2.2 play a crucial role to transfer some surface species to the gas phase.
During the warm up phase, temperature reaches around $\sim 100$ K and almost all the surface species 
are transfered to the gas phase. As a result, we see a sudden rise in the gas phase abundance 
profile at the cost of sharp drop in the ice phase abundance profile. Beyond $100$ K, production 
of surface species is not possible due to the short residence time of the surface species. 
Thus production would continue only in the gas phase. However, beyond $100$ K, gas phase molecules 
are dissociated by various means. Due to the unavailability of reactants, production of gas phase 
species is also hampered.

The peak abundance of the gas phase species is obtained 
at a time when all the surface species are transfered to the gas phase.
Peak abundances of gas phase $\mathrm{CH_3COOCH_3, \ CH_2DCOOCH_3/CH_3COOCH_2D}$, 
$\mathrm{C_2H_5COOCH_3}$ and $\mathrm{CH_3COCH_2OH}$ are found to be $7.17 \times 10^{-10}, 
\ 6.10 \times 10^{-11}, \ 3.50 \times 10^{-12}, \ 1.82 \times 10^{-12}$ respectively. 
For the ice phase species, peak abundances of $\mathrm{CH_3COOCH_3, \ CH_2DCOOCH_3/ CH_3COOCH_2D}$,  
$\mathrm{C_2H_5COOCH_3}$ and $\mathrm{CH_3COCH_2OH}$ are found to be $1.21 \times 10^{-08}, 
\ 1.42 \times 10^{-09}, \ 7.30 \times 10^{-11}, \ 4.18 \times 10^{-11}$ respectively. 
Final abundances (i.e., after $2 \times 10^6$ years) of gas phase 
$\mathrm{CH_3COOCH_3, \ CH_2DCOOCH_3/CH_3COOCH_2D}$, 
$\mathrm{C_2H_5COOCH_3}$ and $\mathrm{CH_3COCH_2OH}$ are found to be $1.07 \times 10^{-18}, 
\ 2.17 \times 10^{-25}, \ 1.64 \times 10^{-14}, \ 1.08 \times 10^{-18}$ respectively.

Column densities of these species could be calculated
by using the following relation used by Shalabiea \& Greenberg (1994) and Das et al. (2011,2013a):
\begin{equation}
N(A) = n_H x_i R,
\end{equation}
where, $n_H$ is the total hydrogen number density, $x_i$ is the abundance of $i^{th}$ species and $R$
is the path length along the line of sight ($= 1.6 \times 10^{21} A_V )/n_H$).
For the simplicity, we use $A_V=A_{Vmax}=150$ for the computation.
In Table 4, we show column density (peak value obtained in our simulation regime) of these 
species in gas/ice phase along with some observational results. From Table 4, it is interesting to note
that despite a low initial elemental abundance of deuterium ($\sim 10^{-5}$, Linsky et al. 1995) in
the interstellar space, singly deuterated isotopomers of methyl acetate are efficiently producing in the ISM.

\begin{table*}
\addtolength{\tabcolsep}{-4pt}
\centering{
\scriptsize

\caption{Column density of various forms of methyl acetate and its two isomers.}
\begin{tabular}{|c|c|c|c|c|}
\hline
{\bf Species}&{\bf Calculated column density}&{\bf Calculated column density}&{\bf Calculated Column density}&{\bf Column density}\\
&{\bf (ice phase peak value)}&{\bf (gas phase peak value)}&{\bf (gas phase final value)}&{\bf (observation/prediction}\\
&{\bf (in $\mathrm{cm^{-2}}$)}&{\bf (in $\mathrm{cm^{-2}}$)}&{\bf (in $\mathrm{cm^{-2}}$)}&{\bf in gas phase) (in $\mathrm{cm^{-2}}$)}\\
\hline\hline
$\mathrm{CH_3COOCH_3}$&$2.91 \times 10^{15}$&$1.72 \times 10^{14}$&$2.58 \times 10^5$&$(4.2 \pm 0.5) \times {10^{15}}^a$\\
$\mathrm{CH_2DCOOCH_3/CH_3COOCH_2D}$&$3.41 \times 10^{14}$&$1.46\times 10^{13}$&$5.20 \times 10^{-2}$&-\\
$\mathrm{C_2H_5OCHO}$&$1.75 \times 10^{13}$&$8.40 \times 10^{11}$&$3.95 \times 10^9$&$(4.5 \pm 1) \times {10^{14}}^a, 5.4 \times {10^{16}}^b$\\
$\mathrm{CH_2COCH_2OH}$&$1.00 \times 10^{13}$&$4.38\times 10^{11}$&$2.59 \times 10^5$&$5.00 \times {10^{12}}^c$\\
\hline
\multicolumn{5}{|c|}{$^a$ Tercero et al. (2013) (in Orion)}\\
\multicolumn{5}{|c|}{$^b$ Belloche et al. (2009) (in Sgr B2(N))}\\
\multicolumn{5}{|c|}{$^c$ Apponi et al. (2006) (in Sgr B2(N))} \\
\hline
\end{tabular}}
\end{table*}
\begin{figure}
\vskip 3cm
\centering{
\vbox{
  \includegraphics[height=7cm,width=8cm]{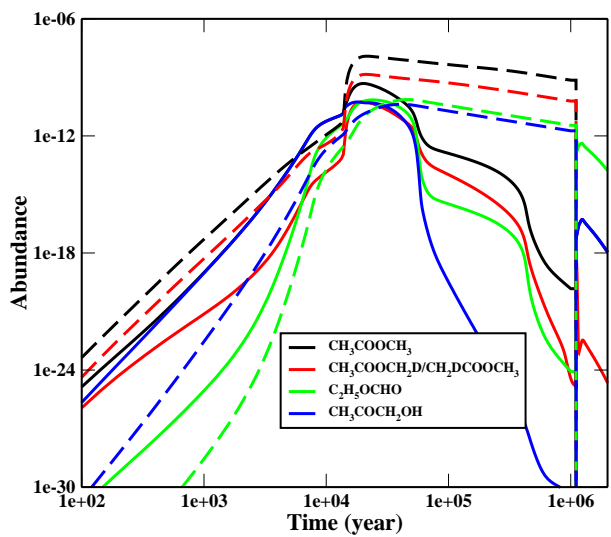}}}
\caption{\small Chemical evolution of methyl acetate and some of its related species. Solid curves represents the gas phase species and dashed curves are used to represent the ice phase species.}
\end{figure}

\section{Spectroscopical modeling and structural calculations}

Past studies reveal that quantum chemical calculations might serve as important
diagnostics for the identification of various species in the ISM. For the computation
of rotational transitions, rotational constants are essential. 
Huang \& Lee (2008) showed that their calculated values often lead to the accuracy within
$20$ MHz for the $B$ and $C$ type constants. 
There are also several instances where calculated vibrational frequencies are accurate to within
$5$ $\mathrm{cm^{-1}}$ or even better \citep{huan08,huan09,huan11,inos11,fort11a,fort11b,fort12a,
fort12b}. Following this type of earlier works, we decided to carry out quantum chemical calculations
to find out spectral properties of various forms of methyl acetate. All our spectroscopic 
calculations do not depend on nuclear spins. This is because ortho and para states  
of a species differ in the arrangement of the nuclear spins and only in cases of NMR shielding and spin-spin 
coupling calculation nuclear spin is used in Gaussian 09W program.  
Other than that, all other properties (energies, geometries, IR, UV-Vis, etc.) 
computed by Gaussian 09W program are nuclear spin independent (as nuclear 
spin does not appear in the Hamiltonian and thus will not change the optimization 
or frequencies). Thus, the values would be identical for different 
ortho and para states and they will be treated as degenerate states for most purposes.

Methyl acetate ($\mathrm{CH_3COOCH_3}$) is an asymmetric top with $C_1$ symmetry. 
Gaussian 09W program is used for the computation of various spectral parameters.
B3LYP \citep{beck93,lee88} functional along with the 6-311G++(d,p) basis set is used for 
the computation of structural parameters as well as ground state energies. 
Ground state energies of $\mathrm{CH_3COOCH_3}$ in gas and 
ice phases (pure water) are found to be $-268.4748$ $a.u$ and $-268.4743$ $a.u.$ respectively. 
These are identical for its deuterated species without zero-point energy 
corrections at $0 \ \mathrm{K}$. The zero point corrections are crucial for the chemistry 
of the deuterated species. Our computed zero point vibrational energies for reactants and 
products of reaction $R1$ along with their deuterated counterpart is shown in Table 5. 

\begin{table*}
\centering{
\scriptsize
\caption{Computed zero-point vibrational energies}
\begin{tabular}{|c|c|c|}
\hline
Species& Zero-point vibrational energy (in gas phase) &  Zero-point vibrational energy (in ice phase)\\
& (in KJ/Mol) & (in KJ/Mol) \\
\hline\hline
CH$_3$CO & 112.845869&  112.65102  \\
CH$_3$O & 97.2549462 &  97.0678377   \\
CH$_2$DCO & 104.789033 &  104.608828    \\
CH$_2$DO & 88.6389605 & 88.4619354   \\
CH$_3$COOCH$_3$ &234.334167  &  233.7121732    \\
CH$_3$COOCH$_2$D & 225.736088&  225.094221     \\
CH$_2$DCOOCH$_3$ & 226.205868 &     225.540403   \\
\hline
\end{tabular}}
\end{table*}

\subsection{Vibrational spectroscopy}

Here also, we use B3LYP functional with 6-311++G(d,p) basis set for 
the computation of the gas phase vibrational transitions of various forms of methyl acetate.
For the ice phase vibrational transitions, Polarizable Continuum Model (PCM) with the integral 
equation formalism variant (IEFPCM) as a default Self-consistent Reaction Field (SCRF) method
are used. 

\begin{figure}
\vskip 1cm
\centering
\includegraphics[height=6cm,width=6cm]{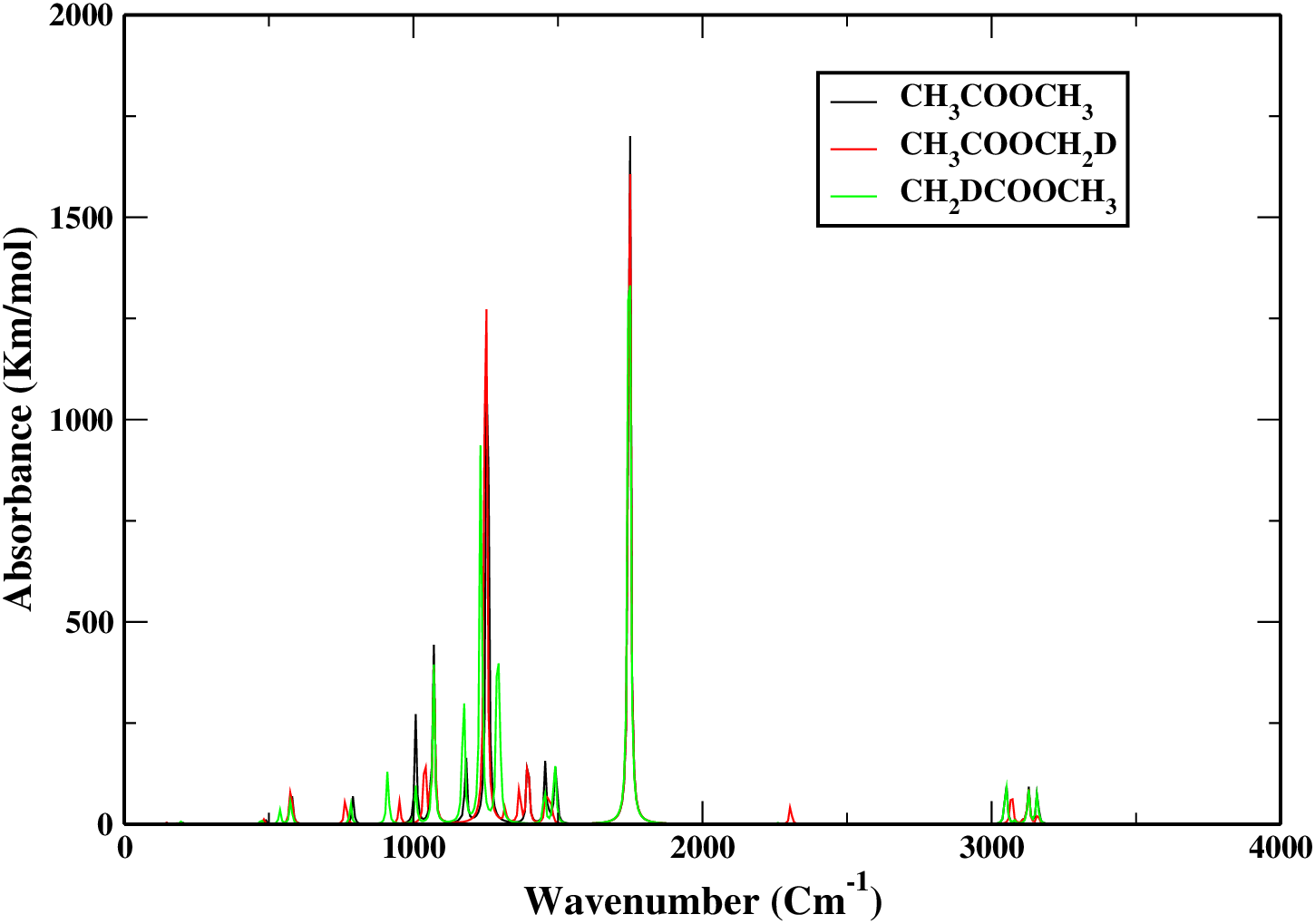}
\caption{Infrared spectra of ice phase $\mathrm{CH_3COOCH_3, \ CH_2DCOOCH3, \ CH3COOCH_2D}$.}
\label{fig-5}
\end{figure}

In Table 6, we present vibrational frequencies of $\mathrm{CH_3COOCH_3}$ and two of its isotopomers 
($\mathrm{CH_2DCOOCH_3}$ and $\mathrm{CH_3COOCH_2D}$). For gas phase 
$\mathrm{CH_3COOCH_3}$, the strongest peak appears at $1261.28 \ \mathrm{cm^{-1}}$. In case of ice phase,
it is noted that this peak appears at $1256.02 \ \mathrm{cm^{-1}}$. Next strongest gas phase peak appears
at $1796.16 \ \mathrm{cm^{-1}}$ which is shifted in the ice phase. In case of 
of various isotopes of methyl acetate, we have different transitions. These differences are attributed 
to the changes of masses of isotopes. Each of $\mathrm{CH_3COOCH_3, \ CH_2DCOOCH_3}$ and $\mathrm{CH_3COOCH_2D}$ 
has a different spectrum. We find that the most intense mode of $\mathrm{CH_2DCOOCH_3}$ in the gas phase  
appears at $1795.20 \ \mathrm{cm^{-1}}$. This peak is shifted in the ice phase by 
nearly $50 \ \mathrm{cm^{-1}}$, i.e., at $1746.56 \ \mathrm{cm^{-1}}$.
The second strongest peak in the gas phase which appears at $1238.04 \ \mathrm{cm^{-1}}$
is also shifted in the ice phase and appears at $1233.76 \ \mathrm{cm^{-1}}$.
In Fig. 3, we show how the isotopic substitution ($\mathrm{CH_2DCOOCH_3, \ CH_3COOCH_2D}$ ) 
plays a part in vibrational transitions of $\mathrm{CH_3COOCH_3}$ in ice phase.
Table 6 clearly shows the differences between spectroscopic parameters 
computed for interstellar methyl acetate in gas phase and in ice phase. 
Moreover, in Table 6, we show band assignments and compare our results with the
existing theoretical and experimental \citep{siva13,geor74} results \citep{sene13}. 
Our results are in good agreement with experimental results on solid phase methyl acetate \citep{siva13}.
Any discrepancy could be attributed to two reasons. First, the experiment was 
performed at $15\ \mathrm{K}$, while we calculate vibrational transitions 
by ab-initio method (i.e., at $0 \ \mathrm{K}$). Second, in our quantum chemical simulation, 
we consider a single methyl acetate inside a spherical cavity while in the experiment, deposited 
numbers of methyl acetate could form clusters. In order to determine column density (by using 
Lambert-Beer relationship used in Bennett et al., 2004) of methyl acetate, it is required 
to find out integrated numerous absorption features. To know the integrated absorption 
features, corresponding integral absorption coefficients (`A' values in
units of $\mathrm{cm \ molecule^{-1}}$) have to be calculated. In Table 6, we show our calculated 
integral absorption coefficients for those transitions, which were present in \citep{siva13}. These
values could be used to predict the column densities of the experimental methyl acetate ice. 


\begin{table*}
\scriptsize{
\centering
\vbox{
\caption{Vibrational frequencies of various forms of methyl acetate in gas phase, in water 
ice at B3LYP/6-311G++(d,p) level of theory}
\begin{tabular}{|c|c|c|c|c|c|c|c|c|}
\hline
\hline
{\bf Species}&{\bf Peak }&{\bf Absorbance}&{\bf Peak }&{\bf Absorbance (in km/mol)/}&{\bf Band}&{\bf Experimental/}\\
{}&{\bf positions }&{{\bf (in km/mol)}}&{\bf positions }&{\bf Integral absorption coefficient }&{\bf assignments}&{\bf calculated}\\

{}&{\bf (Gas phase)}&{}&{\bf (H$_2$O ice)}&{\bf for experimentally}&&{\bf values}\\
&\bf (in $\mathrm{cm^{-1}}$)&& {\bf( in $\mathrm{cm^{-1}}$})&{\bf obtained transitions} &&{\bf (in $\mathrm{cm^{-1}}$)}\\
&&&& (\bf in $\mathrm{cm \ molecule^{-1}}$)&& \\
\hline
&  47.27    & 0.2090       & 54.65     & 0.0002    &$CH_3$ torsion&$43^c$\\
& 130.25 & 0.644  & 146.91 & 1.078  & $CH_3$ torsion&$154^c$\\ 
& 181.78 & 7.76 &  200.71   & 1.421  & torsion&$203^a$,$182^c$\\
& 280.99 & 13.083 & 312.14  & 0.076 & COC bending&$302^a$,$291^c$\\
& 424.03 & 5.595 & 487.62  & 3.774 & CCO bending&$433^a$,$426^c$\\
& 610.49 & 5.636 & 575.95  &  20.921 & C=O wagging &$609^a$,$611^c$\\
& 643.41 & 5.057 & 580.98  & 11.338 & skeletal deformation&$638^a$,$651^c$\\
& 855.46 & 23.369 & 791.11  & 19.998 /$3.32 \times 10^{-18}$ & skeletal deformation&$846^a$, $849^b$, $869^c$\\
& 983.30 & 1.773 & 1008.14   & 78.214 /$1.29 \times 10^{-17}$ & $CH_3$ rocking&$982^a$, $977^b$, $999^c$\\
& 1066.92 & 6.90 & 1058.74  & 14.958 & $CH_3$ rocking&$1073^c$\\
& 1068.53 &  83.157 & 1070.45  & 128.917 /$2.14 \times 10^{-17}$& $O-CH_3$ stretching&$1051^a$,$1044^b$, $1096^c$\\
& 1171.58 &  0.897 &  1160.62 & 1.186 /$1.96 \times 10^{-19}$ & $CH_3$ rocking &$1161^a$,$1191.7^b$, $1192^c$\\
& 1206.93 & 0.737 & 1181.41   &  53.014 & $CH_3$ rocking&$1190^a$,$1219^c$\\
& 1261.28 & 349.202 & 1256.02  &  499.628 /$8.29 \times 10^{-17}$& skeletal deformation&$1249^a$,$1246.2^b$, $1291^c$\\
& 1399.64 & 44.351 & 1396.14  & 63.908 /$1.06 \times 10^{-17}$& $CH_3$ bending&$1372^a$,$1368.7^b$, $1406^c$\\
& 1469.96 & 12.398 &  1456.57 & 44.4885 /$7.38 \times 10^{-18}$& $CH_3$ bending&$1438^a$, $1439.5^b$, $1484^c$\\
{\bf CH$_3$COOCH$_3$}& 1473.17 & 20.3577 & 1468.48 & 4.806 /$7.97 \times 10^{-19}$ & $CH_3$ bending&$1438^a$, $1464.5^b$, $1490^c$\\
& 1477.55 & 8.18 &  1477.07   & 0.8812  & $CH_3$ bending&$1445^a$,$1496^c$\\  
&  1483.88 &  10.634   & 1489.38   &  26.002& $CH_3$ bending &$1462^a$,$1506^c$\\
& 1498.32 & 9.488 & 1494.58  &  31.576& $CH_3$ bending&$1455^a$, $1521^c$\\
& 1796.16  & 288.670  & 1747.67 & 659.267 /$1.09 \times 10^{-16}$& $C=O$ stretching &$1747^a$, $1735.7^b$, $1809^c$\\
& 3050.07  & 6.1048 & 3048.63  & 21.887 /$3.63 \times 10^{-18}$& $CH_3$ stretching&$2955^a$, $2997.7^b$, $3097^c$\\
& 3051.09  &  16.3520  & 3052.37  &  14.406 /$2.39 \times 10^{-18}$&  $CH_3$ stretching&$2940^a$,$3021.5^b$, $3102^c$\\
& 3110.74 &  5.393  & 3110.28 & 2.562& $CH_3$ stretching&$3002^a$,$3186^c$\\
& 3122.45 & 20.665 & 3127.81  & 28.399& $CH_3$ stretching&$3002^a$,$3189^c$\\
& 3155.83 & 9.417 & 3158.73  & 23.852& $CH_3$ stretching&$3028^a$,$3222^c$\\
& 3156.53 & 11.217 & 3160.55  & 1.808& $CH_3$ stretching&$3028^a$,$3225^c$\\
\hline
&  41.04    & 0.067       & 51.68     & 0.0911    & $CH_3$ torsion&\\
& 130.06 & 0.690  & 140.99 & 0.560 & $CH_3$ torsion&\\
& 179.61 & 7.740 &  194.88   & 2.040&   torsion&\\
& 277.79 & 12.446 & 306.38  & 0.124 &COC bending&\\
& 411.93 & 5.797 & 473.08  & 3.566 &CCO bending&\\
& 565.46 & 5.322 & 538.17  &  10.404 & C=O wagging&\\
& 639.25 & 5.150 & 575.97  & 19.820 & skeletal deformation&\\
& 853.19 & 23.930 & 783.20  & 14.752 & skeletal deformation&\\
& 898.09 & 2.462 & 911.30   & 41.344 &  $CH_3$ rocking&\\
& 1007.10 & 5.250 & 1008.63  & 27.630&  $CH_3$ rocking&\\
& 1056.67 &  65.689 & 1070.05  & 120.315 & $O-CH_3$ stretching&\\
& 1171.58 &  0.8675 &  1160.61 & 1.257  & $CH_3$ rocking &\\
& 1206.05 & 1.783 & 1173.46   &  118.272 & $CH_3$ rocking&\\
& 1238.04 & 260.244 & 1233.76  &  321.961& skeletal deformation&\\
& 1302.53 & 79.1404 & 1291.62  & 185.405 & $CH_3$ bending&\\
& 1311.42 & 76.069 &  1302.97 & 26.988 &$CH_3$ bending&\\
{\bf CH$_2$DCOOCH$_3$}& 1458.38 & 11.527 & 1453.26 & 30.510   &$CH_3$ bending&\\
& 1470.87 & 15.677 &  1477.05   & 0.944   &$CH_3$ bending&\\
&  1483.85 &  9.590   & 1487.44   &  34.191 & $CH_3$ bending&\\
& 1498.32 & 9.599 & 1492.18  &  24.069 & $CH_3$ bending&\\
& 1795.20  & 293.616  & 1746.56 & 670.696  & $C=O$ stretching &\\
& 2263.50  & 1.683 & 2262.84  & 1.310  & $CH_3$ stretching&\\
& 3050.34  &  29.712  & 3049.58  & 35.303  &$CH_3$ stretching&\\
& 3078.54 &  4.062  & 3079.32 & 2.277 &$CH_3$ stretching&\\
& 3122.45 & 20.575 & 3127.68  & 26.847 &$CH_3$ stretching&\\
& 3151.52 & 5.663 & 3155.56  & 11.459 &$CH_3$ stretching&\\
& 3155.87 & 14.781 & 3159.20  & 13.631 &$CH_3$ stretching&\\
\hline
\multicolumn{7}{|c|}{$^a$ George et al. (1974) (gas phase experiment)}\\
\multicolumn{7}{|c|}{$^b$ Sivaraman et al. (2014) (ice phase experiment)}\\
\multicolumn{7}{|c|}{$^c$ Senent et al. (2013) (theoretical calculations for gas phase
by using  MP2/cc-pVTZ level of theory)}\\
\hline
\end{tabular}}}
\end{table*}

\clearpage
{\scriptsize
\centering
\vbox{
\begin{tabular}{|c|c|c|c|c|c|c|c|}
\hline
\hline
{\bf Species}&{\bf Peak }&{\bf Absorbance}&{\bf Peak }&{\bf Absorbance (in km/mol)/}&{\bf Band}&{\bf Experimental/}\\
{}&{\bf positions }&{{\bf (in km/mol)}}&{\bf positions }&{\bf Integral absorption coefficient }&{\bf assignments}&{\bf calculated}\\
{}&{\bf (Gas phase)}&{}&{\bf (H$_2$O ice)}&{\bf for experimentally}&&{\bf values}\\
&\bf (in $cm^{-1}$)&& {\bf( in $cm^{-1}$})&{\bf obtained transitions} &&{\bf (in $cm^{-1}$)}\\
&&&& (\bf in cm molecule$^{-1}$)&& \\
\hline
&  47.24    & 0.220       & 49.46     & 0.0009  & $CH_3$ torsion   & \\
& 117.08 & 6.318  & 146.14 & 1.146  & $CH_3$ torsion&\\
& 176.79 & 7.851 &  182.51   & 0.752  &   torsion &\\
& 272.21 & 12.529 & 302.79  & 0.060 &COC bending&\\
& 422.11 & 5.523 & 482.41  & 3.599 &CCO bending&\\
& 610.20 & 5.577 & 573.70  &  19.981 & C=O wagging&\\
& 640.91 & 5.248 & 580.96  & 11.367 & skeletal deformation&\\
& 795.99 & 13.020 & 765.54  & 22.042  & skeletal deformation &\\
& 981.34 & 3.634 & 952.12   & 16.029   &  $CH_3$ rocking&\\
& 1033.62 & 10.259 & 1039.70  & 63.2145  &  $CH_3$ rocking&\\
& 1066.63 &  6.914 & 1058.65  & 15.014 & $O-CH_3$ stretching&\\
& 1068.93 &  80.445 &  1011.82 & 113.975 & $CH_3$ rocking &\\
& 1119.12 & 0.121 & 1111.58   &  0.0899   &$CH_3$ rocking&\\
& 1260.55 & 339.071 & 1250.04  &  570.059 & $O-CH_3$ stretching&\\
& 1320.11 & 6.727 & 1317.61  & 11.506 & $CH_3$ rocking&\\
& 1361.98 & 13.000 &  1367.32 & 30.669& $CH_3$ rocking&\\
{\bf CH$_3$COOCH$_2$D}& 1401.89 & 60.743 & 1396.14 & 62.187&$CH_3$ bending   &\\
& 1472.26 & 16.343 &  1459.23   & 30.448 & $CH_3$ bending &\\
&  1477.58 &  9.424   & 1473.12   &  19.729&  $CH_3$ bending&\\
& 1495.16 & 7.333 & 1480.91 &  5.211& $CH_3$ bending&\\
& 1795.00  & 288.676  & 1747.40 & 661.5813 &$C=O$ stretching &\\
& 2299.03  & 16.867 & 2304.94  & 14.794  &$CH_3$ stretching&\\
& 3050.81 &  2.367  & 3051.19  & 0.099  &$CH_3$ stretching&\\
& 3074.13 &  20.519  & 3069.56 & 30.683 &$CH_3$ stretching&\\
& 3110.73 & 5.394 & 3110.27  & 2.541 &$CH_3$ stretching&\\
& 3122.22 & 20.407 & 3127.54  & 28.001 &$CH_3$ stretching&\\
& 3156.50 & 6.463 & 3160.18  & 8.432 &$CH_3$ stretching&\\
\hline
\end{tabular}}}
\clearpage

\subsection{Rotational spectroscopy}

Density functional theory (DFT) along with some suitable basis sets are known to accurately
calculate spectral properties of various interstellar molecules \citep{rung84,pule10,piev14}.
Accuracy of computed rotational frequencies 
depends on the choices of the models. Recently, Carles et al. (2013, 2014) 
explained experimental rotational frequencies of 4-methylcyanoallene and 
ethyl cyanoacetylene with high-level quantum chemical calculations at the 
B3LYP/cc-pVTZ level of theory by using Gaussian 09W program. Their results are in 
excellent agreement with the experiment. By following their procedure, 
we similarly use Becke’s three parameter hybrid functional \citep{beck88} employing 
the Lee, Yang, and Parr correlation functional (B3LYP; Lee et al., 1988) in  
DFT to calculate rotational constants 
and Watson’s S-reduction quartic and sextic centrifugal distortion constants \citep{wats77}
observing the precautions of McKean et al. (2008). But there are also few other 
studies where CCSD level of theory have been used for the 
computation of rotational and distortional constants \citep{fort12a,fort12b,fort13}. 
Computation of the anharmonic frequencies require the use of analytic second 
derivative of energies at the displaced geometries. But the CCSD method in 
Gaussian 09W program only implements energies. Thus, analytical second derivative of energies 
are not available at this level of theory. In fact, there are no options in Gaussian 
09W program to compute rotational and distortional constants at the CCSD level 
of theory. This prompted us to use B3LYP functional along with the Peterson and Dunning’s 
correlation consistent (cc-pVTZ) basis set \citep{pete02,carl13} which are proven to
yield satisfactory results \citep{carl13,carl14,das15} and also takes less CPU time than 
other highly correlated methods. 
Many molecules have been detected in interstellar space with internal 
rotations \citep{nguy14}. Most of these detection were based on the intense laboratory 
and theoretical work followed by observations with the help of radio telescopes. 
One of the simplest example of such observation is detection of methanol in Orion A 
by Lovas et al., (1976). So as like the methanol which includes one methyl top, 
internal rotation is also important for methyl acetate molecule having two methyl top . 
Nguyen et al., (2014) mentioned that the main problem of analyzing the rotational 
spectrum of methyl acetate is the internal rotation of two inequivalent methyl groups 
with one low ($102.413(20) \ cm^{-1}$) and one intermediate barrier ($424.580(56) \ cm^{-1}$). 
They predicted lines up to $J=30$ using BELGI-Cs-2tops code and by considering ground state 
parameters from Tudoire et al., (2011). However in the experiment of Tudoire et al., (2011), 
laboratory measurements were limited to $J=19$ which includes some uncertainty in the 
predicted lines of Nguyen et al., (2014). In this paper, we have tried to extend the 
line list till $J=41$ by computing many $J$, $K$ dependent parameters as well as the sextic 
centrifugal distortion constants which was not derived in the experiment of Tudoire et al., (2011). 
Our computed parameters considers various symmetries that arise from the interaction of 
overall rotation and the internal rotations of two methyl tops as well as the 
interaction of the methyl tops with each other. We have followed the same technique to extend
the line list for two isotopomers of $\mathrm{CH_3COOCH_3}$.

Computed rotational and vibrational interaction parameters are shown 
in Table 7. Using these rotational and distortional constants, we calculate various rotational transitions
for $\mathrm{CH_3COOCH_2D}$ and $\mathrm{CH_2DCOOCH_3}$ in JPL format (Table 8 and Table 9 respectively)
by using the `SPCAT' program \citep{pick91}. 
These catalog files could be used for the detection of the singly deuterated isotopomers of
methyl acetate in the ISM.

Das et al. (2015) already discussed the importance of theoretical calculations prior to any 
astronomical survey. Our computed frequencies, abundances, column densities for various forms of 
methyl acetate could be used for radiative transfer modeling. These informations would be
crucial for observing any species in the ISM (Coutens et al. 2014 and references therein). CASSIS software 
(an interactive spectrum analyzer, http://cassis.cesr.fr) would be appropriate for modeling
observations from this type of molecules. Since ethyl formate and hydroxy acetone 
were already observed, they were already enlisted within CASSIS.   
Thus we carry out Local Thermodynamic Equilibrium (LTE) modeling for ethyl formate and
hydroxy acetone by using CASSIS software.
Fig. 4a and Fig. 4b shows emission spectra of ethyl formate and hydroxyacetone respectively.
We model this spectrum by using CASSIS software in LTE (assuming the transitions are optically thin). 
For this modeling, we use source size $= 3^"$, excitation temperature ($T_{ex}) = 100 \ \mathrm{K}$ 
(typical hot core condition), line width (FWHM) 
$=7 \ \mathrm{km \ s ^{-1}}$, $V_{lsr}= 64$ $\mathrm{km \ s^{-1}}$, 
column density of Ethyl formate $=5.4 \times 10^{16}$ $\mathrm{cm^{-2}}$ (column density of hydroxyacetone 
$=5.4 \times 10^{16}$ $\mathrm{cm^{-2}}$), column density of $H_2$ $=  1.8 \times 10^{25} \ \mathrm{cm^{-2}}$.
These parameters are used by following the best fit LTE model of ethyl formate from Belloche et al. (2009).
Similar type of modeling would be possible if one use our catalog files (Table 8 and Table 9) under
CASSIS software. Because at present it is not possible to include our catalog files directly into CASSIS,
in Fig. 5, stick diagram for the rotational transitions of methyl acetate is shown by using
ASCP program \citep{kisi98,kisi00}. Here, we use the catalog file generated by the
SPCAT program as an input of ASCP program. Since the partition function in SPACT program 
is calculated at $T=300$ K, Fig. 5 represents the stick diagram at $T=300$ K. 

\begin{figure}
\vskip 1cm
\centering{
\vbox{
  \includegraphics[height=7cm,width=8cm]{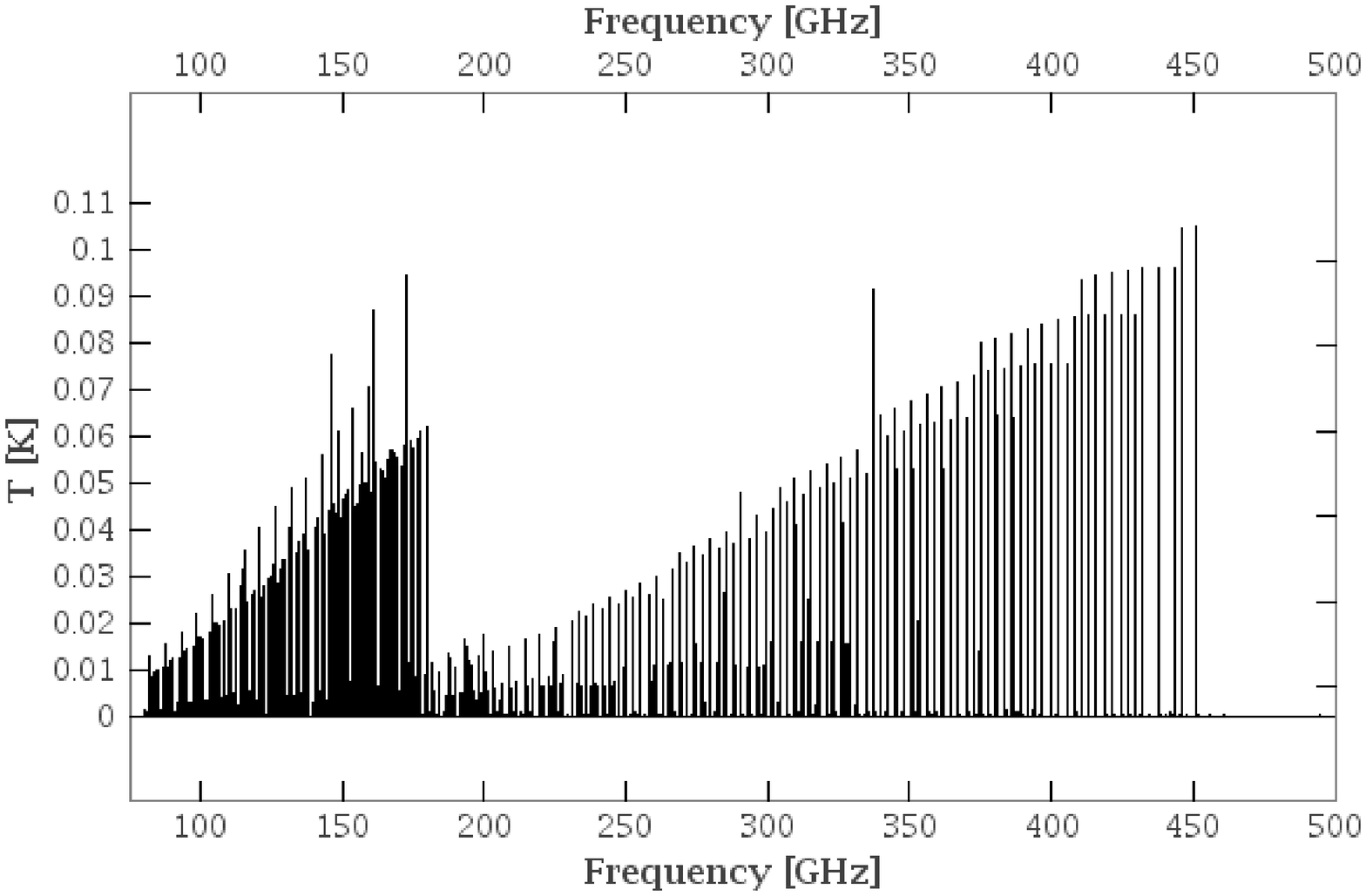}
  \includegraphics[height=7cm,width=8cm]{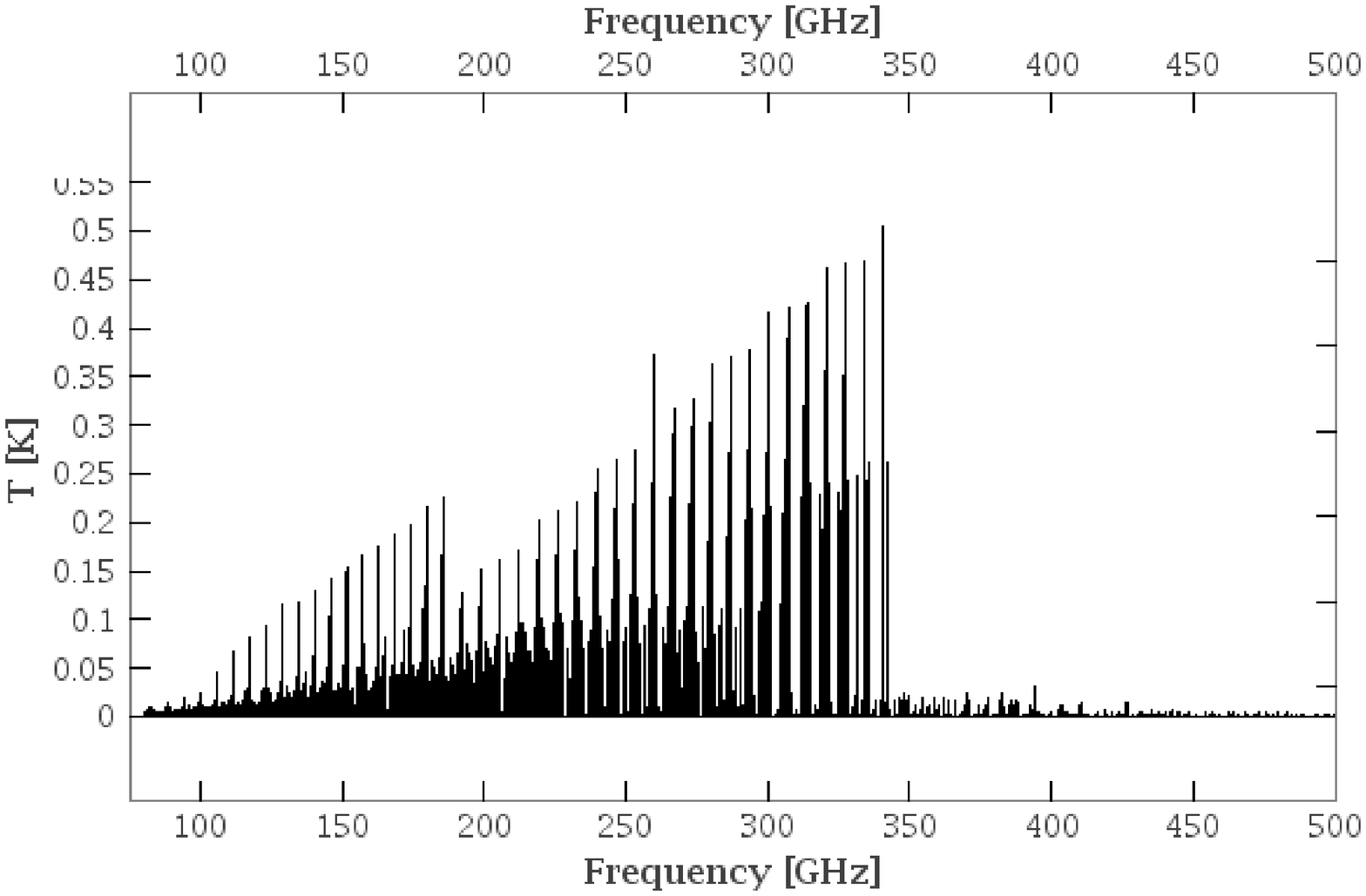}}}
\caption{\small LTE model of (a) Ethyl formate and (b) hydroxyacetone.}
\end{figure}

\begin{figure}
\vskip 1cm
\centering
\includegraphics[height=6cm,width=6cm]{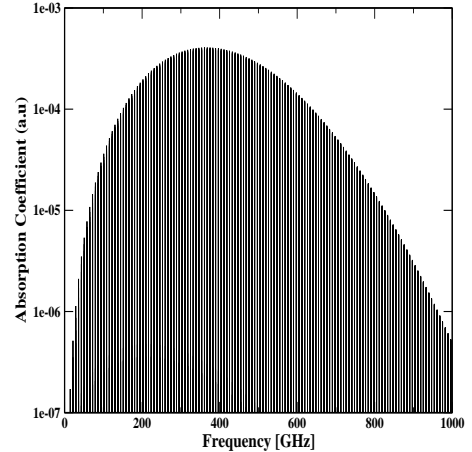}
\caption{Stick diagram of methyl acetate at T=300K}
\label{fig-6}
\end{figure}


\begin{table*}
\scriptsize{
\centering
\vbox{
\caption{Theoretical rotational parameters of methyl acetate and its two 
isotopomers at B3LYP/aug-cc-pVTZ level of theory
}
\begin{tabular}{|c|c|c|c|c|}
\hline
{\bf Species}&{\bf Rotational }&{\bf Values}& 
{\bf Distortional }&{\bf Values} \\
&{\bf constants }&{\bf in $\mathrm{MHz}$}& {\bf  constants }&{\bf in $\mathrm{MHz}$} \\
&&&& \\
\hline
\hline
&A& 10251.95& $D_J$& 0.758$\times$10$^{-3}$ \\
{\bf CH$_3$COOCH$_3$ in gas phase}& B & 4154.13 & $D_{JK}$& 0.440$\times$10$^{-2}$ \\
&C& 3068.82 & $D_{K}$& 0.543$\times$10$^{-2}$    \\
& & & $d_1$ & $-0.186\times$10$^{-3}$ \\
&&& $d_2$& $0.202\times$10$^{-4}$\\
\hline
&A& 9801.87& $D_J$& 0.964$\times$10$^{-3}$ \\
{\bf CH$_2$DCOOCH$_3$ in gas phase}& B & 4017.75 & $D_{JK}$& 0.316$\times$10$^{-2}$ \\
&C& 2979.14 & $D_{K}$& 0.727$\times$10$^{-2}$    \\
& & & $d_1$ & $-0.249\times$10$^{-3}$ \\
&&& $d_2$& $0.683\times$10$^{-4}$\\
\hline
&A& 10090.54& $D_J$& 0.649$\times$10$^{-3}$ \\
{\bf CH$_3$COOCH$_2$D in gas phase}& B & 3935.64 & $D_{JK}$& 0.419$\times$10$^{-2}$ \\
&C& 2934.42 & $D_{K}$& 0.611$\times$10$^{-2}$    \\
& & & $d_1$ & $-0.155\times$10$^{-3}$ \\
&&& $d_2$& $0.144\times$10$^{-4}$\\
\hline
\end{tabular}}}
\end{table*}


\section{Conclusions}

Followings are the major results of this paper:

\noindent {$\bullet$ A complete reaction network is presented to include the formation and 
destruction of methyl acetate along with its two singly deuterated isoropomers and 
two isomers of methyl acetate (ethyl formate and hydroxyacetone). 
Rate coefficients are also calculated indigenously in the absence of any 
available educated estimations. 
}

\noindent{$\bullet$ In order to mimic realistic interstellar features, we use two phase model for the
physical properties of the cloud. In the first phase, the cloud is
collapsing and in the second phase, the cloud is warming up. We couple our
chemical model with this dynamic physical condition and found that 
methyl acetate along with its two singly deuterated isotopomers could be efficiently formed in gas/ice phase.}

\noindent {$\bullet$ Vibrational and rotational transitions 
of the various forms of methyl acetate are explored. Our vibrational transitions are 
compared with the existing experimental/theoretical results and it is
found that our results are in good agreement with the experimental results.
Based on our calculations of the rotational and distortional constants, 
we prepare two catalog files for the singly deuterated methyl acetates, 
which could be useful for its future detection around the region, where methyl
acetate has already been observed.}

\section{Acknowledgments}
AD, PG, DS \& SKC are grateful to ISRO respond (Grant No. ISRO/RES/2/372/11-12) 
and DST (Grant No. SB/S2/HEP-021/2013) for partial financial support. 
LM thanks MOES and ERC starting grant (3DICE, grant
agreement 336474) for funding during this work.
Authors also wish to thank Mr. Soumyadip Mondal (RKMRC, Narendrapur) for his help in data reduction 
during his M.Sc project work at Indian Centre for Space Physics. 
We would like to thank the anonymous referee whose valuable
suggestions were extremely helpful for improving the content of our paper.

\clearpage

\begin{table*}
\scriptsize{
\centering
\vbox{\caption{Rotational transitions for gas phase $\mathrm{CH_3COOCH_2D}$}
\begin{tabular}{|c|c|c|c|c|c|c|c|c|}
\hline
{\bf Frequency$^a$ } &{\bf I$^c$} & {\bf D$^d$ } & {\bf E$_{lower}$$^e $} & {\bf g$_{up}$$^f$ } &Tag$^g$& {\bf QnF$^h$} & {\bf Upper state } & {\bf Lower state }\\
{\bf (in $\mathrm{MHz}$)}  & & & {\bf $ (in \ \mathrm{cm^{-1}})$} &  &&  & {\bf (Qn$_{up}$$^i$}) & {\bf (Qn$_{lower}$$^j$})\\
\hline
    6870.0333 &-8.3550&3 & -0.0000& 1 &74003&304& 1 0 1 0  &  0 0 0 1  \\
    6870.0524 &-7.6560&3 & -0.0000& 5 &74003&304& 1 0 1 2  &  0 0 0 1  \\
    6870.0651 &-7.8779&3 & -0.0000& 3 &74003&304& 1 0 1 1  &  0 0 0 1  \\
   13740.0730 &-7.5775&3 &  0.2292& 3 &74003&304& 2 0 2 1  &  1 0 1 1  \\
   13740.0857 &-8.7536&3 &  0.2292& 3 &74003&304& 2 0 2 1  &  1 0 1 2  \\
   13740.0932 &-6.8294&3 &  0.2292& 7 &74003&304& 2 0 2 3  &  1 0 1 2  \\
   13740.0942 &-7.1004&3 &  0.2292& 5 &74003&304& 2 0 2 2  &  1 0 1 1  \\
   13740.1047 &-7.4526&3 &  0.2292& 3 &74003&304& 2 0 2 1  &  1 0 1 0  \\
   13740.1069 &-7.5775&3 &  0.2292& 5 &74003&304& 2 0 2 2  &  1 0 1 2  \\
   20610.0850 &-7.4027&3 &  0.6875& 5 &74003&304& 3 0 3 2  &  2 0 2 2  \\
   20610.0986 &-8.9467&3 &  0.6875& 5 &74003&304& 3 0 3 2  &  2 0 2 3  \\
   20610.1035 &-6.3393&3 &  0.6875& 9 &74003&304& 3 0 3 4  &  2 0 2 3  \\
   20610.1040 &-6.4996&3 &  0.6875& 7 &74003&304& 3 0 3 3  &  2 0 2 2  \\
   20610.1061 &-6.6703&3 &  0.6875& 5 &74003&304& 3 0 3 2  &  2 0 2 1  \\
   20610.1176 &-7.4027&3 &  0.6875& 7 &74003&304& 3 0 3 3  &  2 0 2 3  \\
   27480.0511 &-7.2794&3 &  1.3750& 7 &74003&304& 4 0 4 3  &  3 0 3 3  \\
   27480.0689 &-5.9881&3 &  1.3750&11 &74003&304& 4 0 4 5  &  3 0 3 4  \\
   27480.0692 &-6.1033&3 &  1.3750& 9 &74003&304& 4 0 4 4  &  3 0 3 3  \\
   27480.0701 &-6.2214&3 &  1.3750& 7 &74003&304& 4 0 4 3  &  3 0 3 2  \\
   27480.0833 &-7.2794&3 &  1.3750& 9 &74003&304& 4 0 4 4  &  3 0 3 4  \\
   34349.9572 &-7.1846&3 &  2.2916& 9 &74003&304& 5 0 5 4  &  4 0 4 4  \\
   34349.9747 &-5.7141&3 &  2.2916&13 &74003&304& 5 0 5 6  &  4 0 4 5  \\
   34349.9749 &-5.8044&3 &  2.2916&11 &74003&304& 5 0 5 5  &  4 0 4 4  \\
   34349.9754 &-5.8958&3 &  2.2916& 9 &74003&304& 5 0 5 4  &  4 0 4 3  \\
   34349.9893 &-7.1846&3 &  2.2916&11 &74003&304& 5 0 5 5  &  4 0 4 5  \\
   41219.7888 &-7.1081&3 &  3.4374&11 &74003&304& 6 0 6 5  &  5 0 5 5  \\
   41219.8060 &-5.4896&3 &  3.4374&15 &74003&304& 6 0 6 7  &  5 0 5 6  \\
   41219.8061 &-5.5640&3 &  3.4374&13 &74003&304& 6 0 6 6  &  5 0 5 5  \\
   41219.8064 &-5.6389&3 &  3.4374&11 &74003&304& 6 0 6 5  &  5 0 5 4  \\
   41219.8208 &-7.1081&3 &  3.4374&13 &74003&304& 6 0 6 6  &  5 0 5 6  \\
   48089.5309 &-7.0442&3 &  4.8123&13 &74003&304& 7 0 7 6  &  6 0 6 6  \\
   48089.5479 &-5.2997&3 &  4.8123&17 &74003&304& 7 0 7 8  &  6 0 6 7  \\
   48089.5480 &-5.3630&3 &  4.8123&15 &74003&304& 7 0 7 7  &  6 0 6 6  \\
   48089.5483 &-5.4266&3 &  4.8123&13 &74003&304& 7 0 7 6  &  6 0 6 5  \\
   48089.5629 &-7.0442&3 &  4.8123&15 &74003&304& 7 0 7 7  &  6 0 6 7  \\
   54959.1688 &-6.9898&3 &  6.4164&15 &74003&304& 8 0 8 7  &  7 0 7 7  \\
   54959.1857 &-5.1353&3 &  6.4164&19 &74003&304& 8 0 8 9  &  7 0 7 8  \\
   54959.1858 &-5.1905&3 &  6.4164&17 &74003&304& 8 0 8 8  &  7 0 7 7  \\
   54959.1859 &-5.2458&3 &  6.4164&15 &74003&304& 8 0 8 7  &  7 0 7 6  \\
   54959.2007 &-6.9898&3 &  6.4164&17 &74003&304& 8 0 8 8  &  7 0 7 8  \\
   61828.6876 &-6.9427&3 &  8.2497&17 &74003&304& 9 0 9 8  &  8 0 8 8  \\
   61828.7043 &-4.9908&3 &  8.2497&21 &74003&304& 9 0 910  &  8 0 8 9  \\
   61828.7044 &-5.0396&3 &  8.2497&19 &74003&304& 9 0 9 9  &  8 0 8 8  \\
   61828.7045 &-5.0886&3 &  8.2497&17 &74003&304& 9 0 9 8  &  8 0 8 7  \\
   61828.7195 &-6.9427&3 &  8.2497&19 &74003&304& 9 0 9 9  &  8 0 8 9  \\
   68698.0724 &-6.9015&3 & 10.3121&19 &74003&304&10 010 9  &  9 0 9 9  \\
   68698.0890 &-4.8620&3 & 10.3121&23 &74003&304&10 01011  &  9 0 910  \\
   68698.0891 &-4.9059&3 & 10.3121&21 &74003&304&10 01010  &  9 0 9 9  \\
   68698.0892 &-4.9498&3 & 10.3121&19 &74003&304&10 010 9  &  9 0 9 8  \\
   68698.1042 &-6.9015&3 & 10.3121&21 &74003&304&10 01010  &  9 0 910  \\
   75567.3082 &-6.8651&3 & 12.6036&21 &74003&304&11 01110  & 10 01010  \\
   75567.3248 &-4.7461&3 & 12.6036&25 &74003&304&11 01112  & 10 01011  \\
   75567.3249 &-4.7860&3 & 12.6036&23 &74003&304&11 01111  & 10 01010  \\
   75567.3250 &-4.8258&3 & 12.6036&21 &74003&304&11 01110  & 10 010 9  \\
   75567.3401 &-6.8651&3 & 12.6036&23 &74003&304&11 01111  & 10 01011  \\
   82436.3804 &-6.8328&3 & 15.1242&23 &74003&304&12 01211  & 11 01111  \\
   82436.3969 &-4.6411&3 & 15.1242&27 &74003&304&12 01213  & 11 01112  \\
   82436.3970 &-4.7140&3 & 15.1242&23 &74003&304&12 01211  & 11 01110  \\
   82436.3970 &-4.6775&3 & 15.1242&25 &74003&304&12 01212  & 11 01111  \\
   82436.4122 &-6.8328&3 & 15.1242&25 &74003&304&12 01212  & 11 01112  \\
   89305.2739 &-6.8041&3 & 17.8740&25 &74003&304&13 01312  & 12 01212  \\
   89305.2904 &-4.5787&3 & 17.8740&27 &74003&304&13 01313  & 12 01212  \\
   89305.2904 &-4.5451&3 & 17.8740&29 &74003&304&13 01314  & 12 01213  \\
   89305.2905 &-4.6124&3 & 17.8740&25 &74003&304&13 01312  & 12 01211  \\
   89305.3057 &-6.8041&3 & 17.8740&27 &74003&304&13 01313  & 12 01213  \\
   96173.9739 &-6.7783&3 & 20.8529&27 &74003&304&14 01413  & 13 01313  \\
   96173.9903 &-4.4883&3 & 20.8529&29 &74003&304&14 01414  & 13 01313  \\
   96173.9903 &-4.4571&3 & 20.8529&31 &74003&304&14 01415  & 13 01314  \\
   96173.9904 &-4.5195&3 & 20.8529&27 &74003&304&14 01413  & 13 01312  \\
   96174.0057 &-6.7783&3 & 20.8529&29 &74003&304&14 01414  & 13 01314  \\
\hline
\end{tabular}}}
\end{table*}

\clearpage
\hskip 1cm
{\scriptsize
\centering
\begin{tabular}{|c|c|c|c|c|c|c|c|c|}
\hline
{\bf Frequency$^a$ } &{\bf I$^c$} & {\bf D$^d$ } & {\bf E$_{lower}$$^e $} & {\bf g$_{up}$$^f$ } &Tag$^g$& {\bf QnF$^h$} & {\bf Upper state } & {\bf Lower state }\\
{\bf (in $\mathrm{MHz}$)}  & & & {\bf $ (in \ \mathrm{cm^{-1})}$} &  &&  & {\bf (Qn$_{up}$$^i$}) & {\bf (Qn$_{lower}$$^j$})\\
\hline
  103042.4654 &-6.7553&3 & 24.0609&29 &74003&304&15 01514  & 14 01414  \\
  103042.4819 &-4.4341&3 & 24.0609&29 &74003&304&15 01514  & 14 01413  \\
  103042.4819 &-4.4050&3 & 24.0609&31 &74003&304&15 01515  & 14 01414  \\
  103042.4819 &-4.3760&3 & 24.0609&33 &74003&304&15 01516  & 14 01415  \\
  103042.4973 &-6.7553&3 & 24.0609&31 &74003&304&15 01515  & 14 01415  \\
  109910.7337 &-6.7347&3 & 27.4981&31 &74003&304&16 01615  & 15 01515  \\
  109910.7501 &-4.3281&3 & 27.4981&33 &74003&304&16 01616  & 15 01515  \\
  109910.7501 &-4.3009&3 & 27.4981&35 &74003&304&16 01617  & 15 01516  \\
  109910.7502 &-4.3554&3 & 27.4981&31 &74003&304&16 01615  & 15 01514  \\
  109910.7655 &-6.7347&3 & 27.4981&33 &74003&304&16 01616  & 15 01516  \\
  116778.7639 &-6.7162&3 & 31.1643&33 &74003&304&17 01716  & 16 01616  \\
  116778.7802 &-4.2568&3 & 31.1643&35 &74003&304&17 01717  & 16 01616  \\
  116778.7802 &-4.2312&3 & 31.1643&37 &74003&304&17 01718  & 16 01617  \\
  116778.7803 &-4.2825&3 & 31.1643&33 &74003&304&17 01716  & 16 01615  \\
  116778.7957 &-6.7162&3 & 31.1643&35 &74003&304&17 01717  & 16 01617  \\
  123646.5409 &-6.6998&3 & 35.0596&35 &74003&304&18 01817  & 17 01717  \\
  123646.5572 &-4.1664&3 & 35.0596&39 &74003&304&18 01819  & 17 01718  \\
  123646.5573 &-4.2148&3 & 35.0596&35 &74003&304&18 01817  & 17 01716  \\
  123646.5573 &-4.1906&3 & 35.0596&37 &74003&304&18 01818  & 17 01717  \\
  123646.5727 &-6.6998&3 & 35.0596&37 &74003&304&18 01818  & 17 01718  \\
  130514.0500 &-6.6851&3 & 39.1840&37 &74003&304&19 01918  & 18 01818  \\
  130514.0663 &-4.1288&3 & 39.1840&39 &74003&304&19 01919  & 18 01818  \\
  130514.0663 &-4.1059&3 & 39.1840&41 &74003&304&19 01920  & 18 01819  \\
  130514.0664 &-4.1517&3 & 39.1840&37 &74003&304&19 01918  & 18 01817  \\
  130514.0818 &-6.6851&3 & 39.1840&39 &74003&304&19 01919  & 18 01819  \\
  137381.2763 &-6.6722&3 & 43.5375&39 &74003&304&20 02019  & 19 01919  \\
  137381.2926 &-4.0930&3 & 43.5375&39 &74003&304&20 02019  & 19 01918  \\
  137381.2926 &-4.0712&3 & 43.5375&41 &74003&304&20 02020  & 19 01919  \\
  137381.2926 &-4.0494&3 & 43.5375&43 &74003&304&20 02021  & 19 01920  \\
  137381.3081 &-6.6722&3 & 43.5375&41 &74003&304&20 02020  & 19 01920  \\
  144248.2049 &-6.6608&3 & 48.1200&41 &74003&304&21 02120  & 20 02020  \\
  144248.2211 &-3.9966&3 & 48.1200&45 &74003&304&21 02122  & 20 02021  \\
  144248.2212 &-4.0380&3 & 48.1200&41 &74003&304&21 02120  & 20 02019  \\
  144248.2212 &-4.0173&3 & 48.1200&43 &74003&304&21 02121  & 20 02020  \\
  144248.2367 &-6.6608&3 & 48.1200&43 &74003&304&21 02121  & 20 02021  \\
  151114.8209 &-6.6508&3 & 52.9316&43 &74003&304&22 02221  & 21 02121  \\
  151114.8371 &-3.9867&3 & 52.9316&43 &74003&304&22 02221  & 21 02120  \\
  151114.8371 &-3.9669&3 & 52.9316&45 &74003&304&22 02222  & 21 02121  \\
  151114.8371 &-3.9471&3 & 52.9316&47 &74003&304&22 02223  & 21 02122  \\
  151114.8527 &-6.6508&3 & 52.9316&45 &74003&304&22 02222  & 21 02122  \\
  157981.1094 &-6.6423&3 & 57.9723&45 &74003&304&23 02322  & 22 02222  \\
  157981.1256 &-3.9386&3 & 57.9723&45 &74003&304&23 02322  & 22 02221  \\
  157981.1256 &-3.9196&3 & 57.9723&47 &74003&304&23 02323  & 22 02222  \\
  157981.1256 &-3.9007&3 & 57.9723&49 &74003&304&23 02324  & 22 02223  \\
  157981.1411 &-6.6423&3 & 57.9723&47 &74003&304&23 02323  & 22 02223  \\
  164847.0555 &-6.6350&3 & 63.2420&47 &74003&304&24 02423  & 23 02323  \\
  164847.0717 &-3.8935&3 & 63.2420&47 &74003&304&24 02423  & 23 02322  \\
  164847.0717 &-3.8754&3 & 63.2420&49 &74003&304&24 02424  & 23 02323  \\
  164847.0717 &-3.8572&3 & 63.2420&51 &74003&304&24 02425  & 23 02324  \\
  164847.0873 &-6.6350&3 & 63.2420&49 &74003&304&24 02424  & 23 02324  \\
  171712.6444 &-6.6290&3 & 68.7407&49 &74003&304&25 02524  & 24 02424  \\
  171712.6606 &-3.8512&3 & 68.7407&49 &74003&304&25 02524  & 24 02423  \\
  171712.6606 &-3.8338&3 & 68.7407&51 &74003&304&25 02525  & 24 02424  \\
  171712.6606 &-3.8164&3 & 68.7407&53 &74003&304&25 02526  & 24 02425  \\
  171712.6761 &-6.6290&3 & 68.7407&51 &74003&304&25 02525  & 24 02425  \\
  178577.8611 &-6.6242&3 & 74.4684&51 &74003&304&26 02625  & 25 02525  \\
  178577.8773 &-3.8116&3 & 74.4684&51 &74003&304&26 02625  & 25 02524  \\
  178577.8773 &-3.7949&3 & 74.4684&53 &74003&304&26 02626  & 25 02525  \\
  178577.8773 &-3.7781&3 & 74.4684&55 &74003&304&26 02627  & 25 02526  \\
  178577.8929 &-6.6242&3 & 74.4684&53 &74003&304&26 02626  & 25 02526  \\
  185442.6908 &-6.6204&3 & 80.4251&53 &74003&304&27 02726  & 26 02626  \\
  185442.7070 &-3.7744&3 & 80.4251&53 &74003&304&27 02726  & 26 02625  \\
  185442.7070 &-3.7583&3 & 80.4251&55 &74003&304&27 02727  & 26 02626  \\
  185442.7070 &-3.7422&3 & 80.4251&57 &74003&304&27 02728  & 26 02627  \\
  185442.7226 &-6.6204&3 & 80.4251&55 &74003&304&27 02727  & 26 02627  \\
  192307.1186 &-6.6178&3 & 86.6108&55 &74003&304&28 02827  & 27 02727  \\
  192307.1348 &-3.7395&3 & 86.6108&55 &74003&304&28 02827  & 27 02726  \\
  192307.1348 &-3.7240&3 & 86.6108&57 &74003&304&28 02828  & 27 02727  \\
  192307.1348 &-3.7085&3 & 86.6108&59 &74003&304&28 02829  & 27 02728  \\
  192307.1504 &-6.6178&3 & 86.6108&57 &74003&304&28 02828  & 27 02728  \\
\hline
\end{tabular}}

\clearpage
\hskip 1cm
{\scriptsize
\centering
\begin{tabular}{|c|c|c|c|c|c|c|c|c|}
\hline
{\bf Frequency$^a$ } & {\bf I$^c$} & {\bf D$^d$ } & {\bf E$_{lower}$$^e $} & {\bf g$_{up}$$^f$ } &Tag$^g$& {\bf QnF$^h$} & {\bf Upper state } & {\bf Lower state }\\
{\bf (in $\mathrm{MHz}$)}  & & & {\bf $ (in \ \mathrm{cm^{-1}})$} &  &&  & {\bf (Qn$_{up}$$^i$}) & {\bf (Qn$_{lower}$$^j$})\\
\hline
  199171.1296 &-6.6161&3 & 93.0255&57 &74003&304&29 02928  & 28 02828  \\
  199171.1457 &-3.6769&3 & 93.0255&61 &74003&304&29 02930  & 28 02829  \\
  199171.1458 &-3.7069&3 & 93.0255&57 &74003&304&29 02928  & 28 02827  \\
  199171.1458 &-3.6919&3 & 93.0255&59 &74003&304&29 02929  & 28 02828  \\
  199171.1614 &-6.6161&3 & 93.0255&59 &74003&304&29 02929  & 28 02829  \\
  206034.7089 &-6.6155&3 & 99.6691&59 &74003&304&30 03029  & 29 02929  \\
  206034.7251 &-3.6763&3 & 99.6691&59 &74003&304&30 03029  & 29 02928  \\
  206034.7251 &-3.6618&3 & 99.6691&61 &74003&304&30 03030  & 29 02929  \\
  206034.7251 &-3.6473&3 & 99.6691&63 &74003&304&30 03031  & 29 02930  \\
  206034.7407 &-6.6155&3 & 99.6691&61 &74003&304&30 03030  & 29 02930  \\
  212897.8417 &-6.6158&3 &106.5417&61 &74003&304&31 03130  & 30 03030  \\
  212897.8578 &-3.6476&3 &106.5417&61 &74003&304&31 03130  & 30 03029  \\
  212897.8578 &-3.6336&3 &106.5417&63 &74003&304&31 03131  & 30 03030  \\
  212897.8578 &-3.6196&3 &106.5417&65 &74003&304&31 03132  & 30 03031  \\
  212897.8734 &-6.6158&3 &106.5417&63 &74003&304&31 03131  & 30 03031  \\
  219760.5130 &-6.6171&3 &113.6432&63 &74003&304&32 03231  & 31 03131  \\
  219760.5291 &-3.6208&3 &113.6432&63 &74003&304&32 03231  & 31 03130  \\
  219760.5291 &-3.6072&3 &113.6432&65 &74003&304&32 03232  & 31 03131  \\
  219760.5291 &-3.5936&3 &113.6432&67 &74003&304&32 03233  & 31 03132  \\
  219760.5447 &-6.6171&3 &113.6432&65 &74003&304&32 03232  & 31 03132  \\
  226622.7079 &-6.6193&3 &120.9736&65 &74003&304&33 03332  & 32 03232  \\
  226622.7240 &-3.5826&3 &120.9736&67 &74003&304&33 03333  & 32 03232  \\
  226622.7240 &-3.5695&3 &120.9736&69 &74003&304&33 03334  & 32 03233  \\
  226622.7241 &-3.5958&3 &120.9736&65 &74003&304&33 03332  & 32 03231  \\
  226622.7397 &-6.6193&3 &120.9736&67 &74003&304&33 03333  & 32 03233  \\
  233484.4117 &-6.6223&3 &128.5330&67 &74003&304&34 03433  & 33 03333  \\
  233484.4278 &-3.5725&3 &128.5330&67 &74003&304&34 03433  & 33 03332  \\
  233484.4278 &-3.5597&3 &128.5330&69 &74003&304&34 03434  & 33 03333  \\
  233484.4278 &-3.5469&3 &128.5330&71 &74003&304&34 03435  & 33 03334  \\
  233484.4434 &-6.6223&3 &128.5330&69 &74003&304&34 03434  & 33 03334  \\
  240345.6093 &-6.6262&3 &136.3212&69 &74003&304&35 03534  & 34 03434  \\
  240345.6254 &-3.5508&3 &136.3212&69 &74003&304&35 03534  & 34 03433  \\
  240345.6254 &-3.5384&3 &136.3212&71 &74003&304&35 03535  & 34 03434  \\
  240345.6254 &-3.5260&3 &136.3212&73 &74003&304&35 03536  & 34 03435  \\
  240345.6411 &-6.6262&3 &136.3212&71 &74003&304&35 03535  & 34 03435  \\
  247206.2859 &-6.6309&3 &144.3382&71 &74003&304&36 03635  & 35 03535  \\
  247206.3020 &-3.5307&3 &144.3382&71 &74003&304&36 03635  & 35 03534  \\
  247206.3020 &-3.5186&3 &144.3382&73 &74003&304&36 03636  & 35 03535  \\
  247206.3020 &-3.5066&3 &144.3382&75 &74003&304&36 03637  & 35 03536  \\
  247206.3177 &-6.6309&3 &144.3382&73 &74003&304&36 03636  & 35 03536  \\
  254066.4267 &-6.6365&3 &152.5841&73 &74003&304&37 03736  & 36 03636  \\
  254066.4428 &-3.5121&3 &152.5841&73 &74003&304&37 03736  & 36 03635  \\
  254066.4428 &-3.5004&3 &152.5841&75 &74003&304&37 03737  & 36 03636  \\
  254066.4428 &-3.4886&3 &152.5841&77 &74003&304&37 03738  & 36 03637  \\
  254066.4584 &-6.6365&3 &152.5841&75 &74003&304&37 03737  & 36 03637  \\
  260926.0167 &-6.6428&3 &161.0589&75 &74003&304&38 03837  & 37 03737  \\
  260926.0327 &-3.4721&3 &161.0589&79 &74003&304&38 03839  & 37 03738  \\
  260926.0328 &-3.4950&3 &161.0589&75 &74003&304&38 03837  & 37 03736  \\
  260926.0328 &-3.4835&3 &161.0589&77 &74003&304&38 03838  & 37 03737  \\
  260926.0484 &-6.6428&3 &161.0589&77 &74003&304&38 03838  & 37 03738  \\
  267785.0410 &-6.6499&3 &169.7624&77 &74003&304&39 03938  & 38 03838  \\
  267785.0571 &-3.4792&3 &169.7624&77 &74003&304&39 03938  & 38 03837  \\
  267785.0571 &-3.4680&3 &169.7624&79 &74003&304&39 03939  & 38 03838  \\
  267785.0571 &-3.4569&3 &169.7624&81 &74003&304&39 03940  & 38 03839  \\
  267785.0728 &-6.6499&3 &169.7624&79 &74003&304&39 03939  & 38 03839  \\
  274643.4848 &-6.6578&3 &178.6948&79 &74003&304&40 04039  & 39 03939  \\
  274643.5009 &-3.4648&3 &178.6948&79 &74003&304&40 04039  & 39 03938  \\
  274643.5009 &-3.4539&3 &178.6948&81 &74003&304&40 04040  & 39 03939  \\
  274643.5009 &-3.4431&3 &178.6948&83 &74003&304&40 04041  & 39 03940  \\
  274643.5166 &-6.6578&3 &178.6948&81 &74003&304&40 04040  & 39 03940  \\
\hline
\end{tabular}}

\clearpage
\hskip 1cm
{\scriptsize
\centering
\begin{tabular}{|c|c|c|c|c|c|c|c|c|}
\hline
{\bf Frequency$^a$ } & {\bf I$^c$} & {\bf D$^d$ } & {\bf E$_{lower}$$^e $} & {\bf g$_{up}$$^f$ } &Tag$^g$& {\bf QnF$^h$} & {\bf Upper state } & {\bf Lower state }\\
{\bf (in $\mathrm{MHz}$)}  & & & {\bf $ (in \ \mathrm{cm^{-1}})$} &  &&  & {\bf (Qn$_{up}$$^i$}) & {\bf (Qn$_{lower}$$^j$})\\
\hline
  199171.1296 &-6.6161&3 & 93.0255&57 &74003&304&29 02928  & 28 02828  \\
  281501.3332 &-6.6664&3 &187.8559&81 &74003&304&41 04140  & 40 04040  \\
  281501.3493 &-3.4517&3 &187.8559&81 &74003&304&41 04140  & 40 04039  \\
  281501.3493 &-3.4411&3 &187.8559&83 &74003&304&41 04141  & 40 04040  \\
  281501.3493 &-3.4305&3 &187.8559&85 &74003&304&41 04142  & 40 04041  \\
  281501.3650 &-6.6664&3 &187.8559&83 &74003&304&41 04141  & 40 04041  \\
  288358.5712 &-6.6757&3 &197.2458&83 &74003&304&42 04241  & 41 04141  \\
  288358.5873 &-3.4398&3 &197.2458&83 &74003&304&42 04241  & 41 04140  \\
  288358.5873 &-3.4295&3 &197.2458&85 &74003&304&42 04242  & 41 04141  \\
  288358.5873 &-3.4191&3 &197.2458&87 &74003&304&42 04243  & 41 04142  \\
  288358.6030 &-6.6757&3 &197.2458&85 &74003&304&42 04242  & 41 04142  \\
  295215.1841 &-6.6858&3 &206.8644&85 &74003&304&43 04342  & 42 04242  \\
  295215.2002 &-3.4292&3 &206.8644&85 &74003&304&43 04342  & 42 04241  \\
  295215.2002 &-3.4191&3 &206.8644&87 &74003&304&43 04343  & 42 04242  \\
  295215.2002 &-3.4090&3 &206.8644&89 &74003&304&43 04344  & 42 04243  \\
  295215.2159 &-6.6858&3 &206.8644&87 &74003&304&43 04343  & 42 04243  \\
\hline
\multicolumn{9}{|c|}{$^a$ Calculated frequency in MHz}\\
\multicolumn{9}{|c|}{in units of MHz then uncertainty of the line is greater or equal to zero.}\\
\multicolumn{9}{|c|}{$^c$ Base 10 logarithm of the integrated intensity at 300K in nm$^2$ MHz}\\
\multicolumn{9}{|c|}{$^d$ Degrees of freedom in the rotational partition function} \\
\multicolumn{9}{|c|}{(0 for atoms, 2 for linear molecules, 3 for non linear molecules)}\\
\multicolumn{9}{|c|}{$^e$ Lower state energy in cm$^{-1}$ relative to the lowest energy} \\
\multicolumn{9}{|c|}{level in the ground vibrionic state.}\\
\multicolumn{9}{|c|}{$^f$ Upper state degeneracy : g$_{up}=g_{I} \times g_{N}$, where g$_{I}$} \\
\multicolumn{9}{|c|}{is the spin statistical weight and g$_{N} =2N+1$ the rotational degeneracy.}\\
\multicolumn{9}{|c|}{$^g$ Molecule Tag}\\
\multicolumn{9}{|c|}{$^h$ Coding for the format of quantum numbers.}\\
\multicolumn{9}{|c|}{QnF=$100 \times Q + 10 \times H + N_{Qn}$; N$_{Qn}$ is the number of quantum }\\
\multicolumn{9}{|c|}{numbers for each state; H indicates the number of half integer quantum numbers;}\\
\multicolumn{9}{|c|}{Qmod5, the residual when Q is divided by 5, gives the number of principal}\\
\multicolumn{9}{|c|}{quantum numbers (without the spin designating ones)}\\
\multicolumn{9}{|c|}{$^i$ Quantum numbers for the upper state}\\
\multicolumn{9}{|c|}{$^j$ Quantum numbers for the lower state}\\
\hline
\end{tabular}}

\clearpage

\begin{table*}
\scriptsize{
\centering
\vbox{\caption{Rotational transitions for gas phase $\mathrm{CH_2DCOOCH_3}$}
\begin{tabular}{|c|c|c|c|c|c|c|c|c|c|}
\hline
{\bf Frequency$^a$ } &{\bf I$^c$} & {\bf D$^d$ } & {\bf E$_{lower}$$^e $} & {\bf g$_{up}$$^f$ } &Tag$^g$& {\bf QnF$^h$} & {\bf Upper state } & {\bf Lower state }\\
{\bf (in $\mathrm{MHz}$)}  & & & {\bf $ (in \ \mathrm{cm^{-1}})$} &  &&  & {\bf (Qn$_{up}$$^i$}) & {\bf (Qn$_{lower}$$^j$})\\
\hline
    6996.8647 &-7.8605&3 & -0.0000& 3 &74003&304& 1 0 1 1  &  0 0 0 1  \\
    6996.8923 &-7.6387&3 & -0.0000& 5 &74003&304& 1 0 1 2  &  0 0 0 1  \\
    6996.9337 &-8.3376&3 & -0.0000& 1 &74003&304& 1 0 1 0  &  0 0 0 1  \\
   13993.7279 &-7.5602&3 &  0.2334& 5 &74003&304& 2 0 2 2  &  1 0 1 2  \\
   13993.7325 &-7.4353&3 &  0.2334& 3 &74003&304& 2 0 2 1  &  1 0 1 0  \\
   13993.7555 &-7.0831&3 &  0.2334& 5 &74003&304& 2 0 2 2  &  1 0 1 1  \\
   13993.7575 &-6.8120&3 &  0.2334& 7 &74003&304& 2 0 2 3  &  1 0 1 2  \\
   13993.7739 &-8.7363&3 &  0.2334& 3 &74003&304& 2 0 2 1  &  1 0 1 2  \\
   13993.8015 &-7.5602&3 &  0.2334& 3 &74003&304& 2 0 2 1  &  1 0 1 1  \\
   20990.5541 &-7.3853&3 &  0.7002& 7 &74003&304& 3 0 3 3  &  2 0 2 3  \\
   20990.5791 &-6.6529&3 &  0.7002& 5 &74003&304& 3 0 3 2  &  2 0 2 1  \\
   20990.5837 &-6.4822&3 &  0.7002& 7 &74003&304& 3 0 3 3  &  2 0 2 2  \\
   20990.5848 &-6.3219&3 &  0.7002& 9 &74003&304& 3 0 3 4  &  2 0 2 3  \\
   20990.5955 &-8.9294&3 &  0.7002& 5 &74003&304& 3 0 3 2  &  2 0 2 3  \\
   20990.6251 &-7.3853&3 &  0.7002& 5 &74003&304& 3 0 3 2  &  2 0 2 2  \\
   27987.3216 &-7.2621&3 &  1.4003& 9 &74003&304& 4 0 4 4  &  3 0 3 4  \\
   27987.3503 &-6.2041&3 &  1.4003& 7 &74003&304& 4 0 4 3  &  3 0 3 2  \\
   27987.3522 &-6.0860&3 &  1.4003& 9 &74003&304& 4 0 4 4  &  3 0 3 3  \\
   27987.3529 &-5.9708&3 &  1.4003&11 &74003&304& 4 0 4 5  &  3 0 3 4  \\
   27987.3917 &-7.2621&3 &  1.4003& 7 &74003&304& 4 0 4 3  &  3 0 3 3  \\
   34984.0100 &-7.1674&3 &  2.3339&11 &74003&304& 5 0 5 5  &  4 0 4 5  \\
   34984.0403 &-5.8786&3 &  2.3339& 9 &74003&304& 5 0 5 4  &  4 0 4 3  \\
   34984.0414 &-5.7872&3 &  2.3339&11 &74003&304& 5 0 5 5  &  4 0 4 4  \\
   34984.0419 &-5.6969&3 &  2.3339&13 &74003&304& 5 0 5 6  &  4 0 4 5  \\
   34984.0797 &-7.1674&3 &  2.3339& 9 &74003&304& 5 0 5 4  &  4 0 4 4  \\
   41980.5995 &-7.0909&3 &  3.5008&13 &74003&304& 6 0 6 6  &  5 0 5 6  \\
   41980.6306 &-5.6217&3 &  3.5008&11 &74003&304& 6 0 6 5  &  5 0 5 4  \\
   41980.6313 &-5.5468&3 &  3.5008&13 &74003&304& 6 0 6 6  &  5 0 5 5  \\
   41980.6317 &-5.4724&3 &  3.5008&15 &74003&304& 6 0 6 7  &  5 0 5 6  \\
   41980.6690 &-7.0909&3 &  3.5008&11 &74003&304& 6 0 6 5  &  5 0 5 5  \\
   48977.0699 &-7.0271&3 &  4.9012&15 &74003&304& 7 0 7 7  &  6 0 6 7  \\
   48977.1016 &-5.4095&3 &  4.9012&13 &74003&304& 7 0 7 6  &  6 0 6 5  \\
   48977.1021 &-5.3459&3 &  4.9012&15 &74003&304& 7 0 7 7  &  6 0 6 6  \\
   48977.1024 &-5.2825&3 &  4.9012&17 &74003&304& 7 0 7 8  &  6 0 6 7  \\
   48977.1393 &-7.0271&3 &  4.9012&13 &74003&304& 7 0 7 6  &  6 0 6 6  \\
   55973.4014 &-6.9728&3 &  6.5349&17 &74003&304& 8 0 8 8  &  7 0 7 8  \\
   55973.4336 &-5.2287&3 &  6.5349&15 &74003&304& 8 0 8 7  &  7 0 7 6  \\
   55973.4339 &-5.1734&3 &  6.5349&17 &74003&304& 8 0 8 8  &  7 0 7 7  \\
   55973.4341 &-5.1183&3 &  6.5349&19 &74003&304& 8 0 8 9  &  7 0 7 8  \\
   55973.4707 &-6.9728&3 &  6.5349&15 &74003&304& 8 0 8 7  &  7 0 7 7  \\
   62969.5742 &-6.9257&3 &  8.4019&19 &74003&304& 9 0 9 9  &  8 0 8 9  \\
   62969.6067 &-5.0716&3 &  8.4019&17 &74003&304& 9 0 9 8  &  8 0 8 7  \\
   62969.6069 &-5.0227&3 &  8.4019&19 &74003&304& 9 0 9 9  &  8 0 8 8  \\
   62969.6071 &-4.9738&3 &  8.4019&21 &74003&304& 9 0 910  &  8 0 8 9  \\
   62969.6435 &-6.9257&3 &  8.4019&17 &74003&304& 9 0 9 8  &  8 0 8 8  \\
   69965.5684 &-6.8846&3 & 10.5024&21 &74003&304&10 01010  &  9 0 910  \\
   69965.6010 &-4.9329&3 & 10.5024&19 &74003&304&10 010 9  &  9 0 9 8  \\
   69965.6013 &-4.8890&3 & 10.5024&21 &74003&304&10 01010  &  9 0 9 9  \\
   69965.6014 &-4.8451&3 & 10.5024&23 &74003&304&10 01011  &  9 0 910  \\
   69965.6376 &-6.8846&3 & 10.5024&19 &74003&304&10 010 9  &  9 0 9 9  \\
   76961.3641 &-6.8483&3 & 12.8362&23 &74003&304&11 01111  & 10 01011  \\
   76961.3969 &-4.8090&3 & 12.8362&21 &74003&304&11 01110  & 10 010 9  \\
   76961.3971 &-4.7691&3 & 12.8362&23 &74003&304&11 01111  & 10 01010  \\
   76961.3972 &-4.7293&3 & 12.8362&25 &74003&304&11 01112  & 10 01011  \\
   76961.4332 &-6.8483&3 & 12.8362&21 &74003&304&11 01110  & 10 01010  \\
   83956.9414 &-6.8161&3 & 15.4033&25 &74003&304&12 01212  & 11 01112  \\
   83956.9744 &-4.6973&3 & 15.4033&23 &74003&304&12 01211  & 11 01110  \\
   83956.9745 &-4.6608&3 & 15.4033&25 &74003&304&12 01212  & 11 01111  \\
   83956.9746 &-4.6244&3 & 15.4033&27 &74003&304&12 01213  & 11 01112  \\
   83957.0105 &-6.8161&3 & 15.4033&23 &74003&304&12 01211  & 11 01111  \\
   90952.2805 &-6.7875&3 & 18.2038&27 &74003&304&13 01313  & 12 01213  \\
   90952.3136 &-4.5958&3 & 18.2038&25 &74003&304&13 01312  & 12 01211  \\
   90952.3137 &-4.5622&3 & 18.2038&27 &74003&304&13 01313  & 12 01212  \\
   90952.3138 &-4.5285&3 & 18.2038&29 &74003&304&13 01314  & 12 01213  \\
   90952.3496 &-6.7875&3 & 18.2038&25 &74003&304&13 01312  & 12 01212  \\
\hline
\end{tabular}}}
\end{table*}

\clearpage
\hskip 1cm
{\scriptsize
\centering
\begin{tabular}{|c|c|c|c|c|c|c|c|c|}
\hline
{\bf Frequency$^a$ } &{\bf I$^c$} & {\bf D$^d$ } & {\bf E$_{lower}$$^e $} & {\bf g$_{up}$$^f$ } &Tag$^g$& {\bf QnF$^h$} & {\bf Upper state } & {\bf Lower state }\\
{\bf (in $\mathrm{MHz}$)}  & & & {\bf $ (in \ \mathrm{cm^{-1}})$} &  &&  & {\bf (Qn$_{up}$$^i$}) & {\bf (Qn$_{lower}$$^j$})\\
\hline
   97947.3615 &-6.7619&3 & 21.2377&29 &74003&304&14 01414  & 13 01314  \\
   97947.3947 &-4.5030&3 & 21.2377&27 &74003&304&14 01413  & 13 01312  \\
   97947.3948 &-4.4718&3 & 21.2377&29 &74003&304&14 01414  & 13 01313  \\
   97947.3949 &-4.4406&3 & 21.2377&31 &74003&304&14 01415  & 13 01314  \\
   97947.4306 &-6.7619&3 & 21.2377&27 &74003&304&14 01413  & 13 01313  \\
  104942.1646 &-6.7389&3 & 24.5049&31 &74003&304&15 01515  & 14 01415  \\
  104942.1979 &-4.4178&3 & 24.5049&29 &74003&304&15 01514  & 14 01413  \\
  104942.1980 &-4.3887&3 & 24.5049&31 &74003&304&15 01515  & 14 01414  \\
  104942.1981 &-4.3596&3 & 24.5049&33 &74003&304&15 01516  & 14 01415  \\
  104942.2337 &-6.7389&3 & 24.5049&29 &74003&304&15 01514  & 14 01414  \\
  111936.6699 &-6.7185&3 & 28.0054&33 &74003&304&16 01616  & 15 01516  \\
  111936.7033 &-4.3392&3 & 28.0054&31 &74003&304&16 01615  & 15 01514  \\
  111936.7034 &-4.3119&3 & 28.0053&33 &74003&304&16 01616  & 15 01515  \\
  111936.7035 &-4.2847&3 & 28.0054&35 &74003&304&16 01617  & 15 01516  \\
  111936.7390 &-6.7185&3 & 28.0053&31 &74003&304&16 01615  & 15 01515  \\
  118930.8576 &-6.7002&3 & 31.7392&35 &74003&304&17 01717  & 16 01617  \\
  118930.8911 &-4.2664&3 & 31.7392&33 &74003&304&17 01716  & 16 01615  \\
  118930.8912 &-4.2408&3 & 31.7392&35 &74003&304&17 01717  & 16 01616  \\
  118930.8912 &-4.2151&3 & 31.7392&37 &74003&304&17 01718  & 16 01617  \\
  118930.9267 &-6.7002&3 & 31.7392&33 &74003&304&17 01716  & 16 01616  \\
  125924.7079 &-6.6839&3 & 35.7063&37 &74003&304&18 01818  & 17 01718  \\
  125924.7414 &-4.1989&3 & 35.7063&35 &74003&304&18 01817  & 17 01716  \\
  125924.7414 &-4.1747&3 & 35.7063&37 &74003&304&18 01818  & 17 01717  \\
  125924.7415 &-4.1505&3 & 35.7063&39 &74003&304&18 01819  & 17 01718  \\
  125924.7769 &-6.6839&3 & 35.7063&35 &74003&304&18 01817  & 17 01717  \\
  132918.2007 &-6.6694&3 & 39.9067&39 &74003&304&19 01919  & 18 01819  \\
  132918.2343 &-4.1360&3 & 39.9067&37 &74003&304&19 01918  & 18 01817  \\
  132918.2344 &-4.1131&3 & 39.9067&39 &74003&304&19 01919  & 18 01818  \\
  132918.2344 &-4.0902&3 & 39.9067&41 &74003&304&19 01920  & 18 01819  \\
  132918.2698 &-6.6694&3 & 39.9067&37 &74003&304&19 01918  & 18 01818  \\
  139911.3164 &-6.6566&3 & 44.3403&41 &74003&304&20 02020  & 19 01920  \\
  139911.3500 &-4.0774&3 & 44.3403&39 &74003&304&20 02019  & 19 01918  \\
  139911.3501 &-4.0556&3 & 44.3403&41 &74003&304&20 02020  & 19 01919  \\
  139911.3501 &-4.0339&3 & 44.3403&43 &74003&304&20 02021  & 19 01920  \\
  139911.3855 &-6.6566&3 & 44.3403&39 &74003&304&20 02019  & 19 01919  \\
  146904.0351 &-6.6454&3 & 49.0073&43 &74003&304&21 02121  & 20 02021  \\
  146904.0687 &-4.0227&3 & 49.0073&41 &74003&304&21 02120  & 20 02019  \\
  146904.0688 &-4.0019&3 & 49.0073&43 &74003&304&21 02121  & 20 02020  \\
  146904.0688 &-3.9812&3 & 49.0073&45 &74003&304&21 02122  & 20 02021  \\
  146904.1041 &-6.6454&3 & 49.0073&41 &74003&304&21 02120  & 20 02020  \\
  153896.3368 &-6.6357&3 & 53.9075&45 &74003&304&22 02222  & 21 02122  \\
  153896.3705 &-3.9715&3 & 53.9075&43 &74003&304&22 02221  & 21 02120  \\
  153896.3705 &-3.9517&3 & 53.9075&45 &74003&304&22 02222  & 21 02121  \\
  153896.3706 &-3.9319&3 & 53.9075&47 &74003&304&22 02223  & 21 02122  \\
  153896.4059 &-6.6357&3 & 53.9075&43 &74003&304&22 02221  & 21 02121  \\
  160888.2018 &-6.6273&3 & 59.0409&47 &74003&304&23 02323  & 22 02223  \\
  160888.2355 &-3.9236&3 & 59.0409&45 &74003&304&23 02322  & 22 02221  \\
  160888.2356 &-3.9047&3 & 59.0409&47 &74003&304&23 02323  & 22 02222  \\
  160888.2356 &-3.8858&3 & 59.0409&49 &74003&304&23 02324  & 22 02223  \\
  160888.2708 &-6.6273&3 & 59.0409&45 &74003&304&23 02322  & 22 02222  \\
  167879.6102 &-6.6203&3 & 64.4076&49 &74003&304&24 02424  & 23 02324  \\
  167879.6440 &-3.8787&3 & 64.4076&47 &74003&304&24 02423  & 23 02322  \\
  167879.6440 &-3.8606&3 & 64.4076&49 &74003&304&24 02424  & 23 02323  \\
  167879.6440 &-3.8425&3 & 64.4076&51 &74003&304&24 02425  & 23 02324  \\
  167879.6792 &-6.6203&3 & 64.4076&47 &74003&304&24 02423  & 23 02323  \\
  174870.5421 &-6.6145&3 & 70.0074&51 &74003&304&25 02525  & 24 02425  \\
  174870.5759 &-3.8367&3 & 70.0074&49 &74003&304&25 02524  & 24 02423  \\
  174870.5760 &-3.8193&3 & 70.0074&51 &74003&304&25 02525  & 24 02424  \\
  174870.5760 &-3.8019&3 & 70.0074&53 &74003&304&25 02526  & 24 02425  \\
  174870.6112 &-6.6145&3 & 70.0074&49 &74003&304&25 02524  & 24 02424  \\
  181860.9777 &-6.6098&3 & 75.8405&53 &74003&304&26 02626  & 25 02526  \\
  181861.0116 &-3.7973&3 & 75.8405&51 &74003&304&26 02625  & 25 02524  \\
  181861.0116 &-3.7805&3 & 75.8405&53 &74003&304&26 02626  & 25 02525  \\
  181861.0116 &-3.7638&3 & 75.8405&55 &74003&304&26 02627  & 25 02526  \\
  181861.0468 &-6.6098&3 & 75.8405&51 &74003&304&26 02625  & 25 02525  \\
  188850.8972 &-6.6063&3 & 81.9067&55 &74003&304&27 02727  & 26 02627  \\
  188850.9311 &-3.7603&3 & 81.9067&53 &74003&304&27 02726  & 26 02625  \\
  188850.9311 &-3.7442&3 & 81.9067&55 &74003&304&27 02727  & 26 02626  \\
  188850.9311 &-3.7281&3 & 81.9067&57 &74003&304&27 02728  & 26 02627  \\
  188850.9662 &-6.6063&3 & 81.9067&53 &74003&304&27 02726  & 26 02626  \\
\hline
\end{tabular}}

\clearpage
\hskip 1cm
{\scriptsize
\centering
\begin{tabular}{|c|c|c|c|c|c|c|c|c|}
\hline
{\bf Frequency$^a$ } &{\bf I$^c$} & {\bf D$^d$ } & {\bf E$_{lower}$$^e $} & {\bf g$_{up}$$^f$ } &Tag$^g$& {\bf QnF$^h$} & {\bf Upper state } & {\bf Lower state }\\
{\bf (in $\mathrm{MHz}$)}  & & & {\bf $ (in \ \mathrm{cm^{-1}})$} &  &&  & {\bf (Qn$_{up}$$^i$}) & {\bf (Qn$_{lower}$$^j$})\\
\hline
  195840.2807 &-6.6039&3 & 88.2061&57 &74003&304&28 02828  & 27 02728  \\
  195840.3145 &-3.7257&3 & 88.2061&55 &74003&304&28 02827  & 27 02726  \\
  195840.3146 &-3.7102&3 & 88.2061&57 &74003&304&28 02828  & 27 02727  \\
  195840.3146 &-3.6946&3 & 88.2061&59 &74003&304&28 02829  & 27 02728  \\
  195840.3497 &-6.6039&3 & 88.2061&55 &74003&304&28 02827  & 27 02727  \\
  202829.1082 &-6.6026&3 & 94.7386&59 &74003&304&29 02929  & 28 02829  \\
  202829.1421 &-3.6933&3 & 94.7386&57 &74003&304&29 02928  & 28 02827  \\
  202829.1422 &-3.6783&3 & 94.7386&59 &74003&304&29 02929  & 28 02828  \\
  202829.1422 &-3.6633&3 & 94.7386&61 &74003&304&29 02930  & 28 02829  \\
  202829.1773 &-6.6026&3 & 94.7386&57 &74003&304&29 02928  & 28 02828  \\
  209817.3601 &-6.6022&3 &101.5043&61 &74003&304&30 03030  & 29 02930  \\
  209817.3940 &-3.6629&3 &101.5043&59 &74003&304&30 03029  & 29 02928  \\
  209817.3940 &-3.6484&3 &101.5043&61 &74003&304&30 03030  & 29 02929  \\
  209817.3941 &-3.6339&3 &101.5043&63 &74003&304&30 03031  & 29 02930  \\
  209817.4291 &-6.6022&3 &101.5043&59 &74003&304&30 03029  & 29 02929  \\
  216805.0164 &-6.6028&3 &108.5030&63 &74003&304&31 03131  & 30 03031  \\
  216805.0503 &-3.6346&3 &108.5030&61 &74003&304&31 03130  & 30 03029  \\
  216805.0504 &-3.6205&3 &108.5030&63 &74003&304&31 03131  & 30 03030  \\
  216805.0504 &-3.6065&3 &108.5030&65 &74003&304&31 03132  & 30 03031  \\
  216805.0854 &-6.6028&3 &108.5030&61 &74003&304&31 03130  & 30 03030  \\
  223792.0573 &-6.6043&3 &115.7349&65 &74003&304&32 03232  & 31 03132  \\
  223792.0912 &-3.6081&3 &115.7349&63 &74003&304&32 03231  & 31 03130  \\
  223792.0912 &-3.5945&3 &115.7349&65 &74003&304&32 03232  & 31 03131  \\
  223792.0913 &-3.5809&3 &115.7349&67 &74003&304&32 03233  & 31 03132  \\
  223792.1263 &-6.6043&3 &115.7349&63 &74003&304&32 03231  & 31 03131  \\
  230778.4629 &-6.6068&3 &123.1998&67 &74003&304&33 03333  & 32 03233  \\
  230778.4968 &-3.5833&3 &123.1998&65 &74003&304&33 03332  & 32 03231  \\
  230778.4968 &-3.5702&3 &123.1998&67 &74003&304&33 03333  & 32 03232  \\
  230778.4969 &-3.5570&3 &123.1998&69 &74003&304&33 03334  & 32 03233  \\
  230778.5319 &-6.6068&3 &123.1998&65 &74003&304&33 03332  & 32 03232  \\
  237764.2133 &-6.6101&3 &130.8977&69 &74003&304&34 03434  & 33 03334  \\
  237764.2473 &-3.5603&3 &130.8977&67 &74003&304&34 03433  & 33 03332  \\
  237764.2473 &-3.5476&3 &130.8977&69 &74003&304&34 03434  & 33 03333  \\
  237764.2473 &-3.5348&3 &130.8977&71 &74003&304&34 03435  & 33 03334  \\
  237764.2823 &-6.6101&3 &130.8977&67 &74003&304&34 03433  & 33 03333  \\
  244749.2888 &-6.6143&3 &138.8287&71 &74003&304&35 03535  & 34 03435  \\
  244749.3228 &-3.5390&3 &138.8287&69 &74003&304&35 03534  & 34 03433  \\
  244749.3228 &-3.5266&3 &138.8287&71 &74003&304&35 03535  & 34 03434  \\
  244749.3228 &-3.5141&3 &138.8287&73 &74003&304&35 03536  & 34 03435  \\
  244749.3578 &-6.6143&3 &138.8287&69 &74003&304&35 03534  & 34 03434  \\
  251733.6694 &-6.6194&3 &146.9926&73 &74003&304&36 03636  & 35 03536  \\
  251733.7035 &-3.5192&3 &146.9926&71 &74003&304&36 03635  & 35 03534  \\
  251733.7035 &-3.5071&3 &146.9926&73 &74003&304&36 03636  & 35 03535  \\
  251733.7035 &-3.4950&3 &146.9926&75 &74003&304&36 03637  & 35 03536  \\
  251733.7385 &-6.6194&3 &146.9926&71 &74003&304&36 03635  & 35 03535  \\
  258717.3354 &-6.6252&3 &155.3896&75 &74003&304&37 03737  & 36 03637  \\
  258717.3694 &-3.5009&3 &155.3896&73 &74003&304&37 03736  & 36 03635  \\
  258717.3694 &-3.4892&3 &155.3896&75 &74003&304&37 03737  & 36 03636  \\
  258717.3695 &-3.4774&3 &155.3896&77 &74003&304&37 03738  & 36 03637  \\
  258717.4044 &-6.6252&3 &155.3896&73 &74003&304&37 03736  & 36 03636  \\
\hline
\end{tabular}}

\clearpage
\hskip 1cm
{\scriptsize
\centering
\begin{tabular}{|c|c|c|c|c|c|c|c|c|}
\hline
{\bf Frequency$^a$ }  &{\bf I$^c$} & {\bf D$^d$ } & {\bf E$_{lower}$$^e $} & {\bf g$_{up}$$^f$ } &Tag$^g$& {\bf QnF$^h$} & {\bf Upper state } & {\bf Lower state }\\
{\bf (in $\mathrm{MHz}$)}  & & & {\bf $ (in \ \mathrm{cm^{-1}})$} &  &&  & {\bf (Qn$_{up}$$^i$}) & {\bf (Qn$_{lower}$$^j$})\\
\hline
  265700.2668 &-6.6319&3 &164.0194&77 &74003&304&38 03838  & 37 03738  \\
  265700.3009 &-3.4841&3 &164.0194&75 &74003&304&38 03837  & 37 03736  \\
  265700.3009 &-3.4726&3 &164.0194&77 &74003&304&38 03838  & 37 03737  \\
  265700.3009 &-3.4612&3 &164.0194&79 &74003&304&38 03839  & 37 03738  \\
  265700.3358 &-6.6319&3 &164.0194&75 &74003&304&38 03837  & 37 03737  \\
  272682.4438 &-6.6394&3 &172.8823&79 &74003&304&39 03939  & 38 03839  \\
  272682.4779 &-3.4687&3 &172.8823&77 &74003&304&39 03938  & 38 03837  \\
  272682.4779 &-3.4575&3 &172.8822&79 &74003&304&39 03939  & 38 03838  \\
  272682.4779 &-3.4464&3 &172.8823&81 &74003&304&39 03940  & 38 03839  \\
  272682.5129 &-6.6394&3 &172.8822&77 &74003&304&39 03938  & 38 03838  \\
  279663.8466 &-6.6476&3 &181.9780&81 &74003&304&40 04040  & 39 03940  \\
  279663.8807 &-3.4546&3 &181.9780&79 &74003&304&40 04039  & 39 03938  \\
  279663.8807 &-3.4437&3 &181.9780&81 &74003&304&40 04040  & 39 03939  \\
  279663.8807 &-3.4329&3 &181.9780&83 &74003&304&40 04041  & 39 03940  \\
  279663.9156 &-6.6476&3 &181.9780&79 &74003&304&40 04039  & 39 03939  \\
  286644.4553 &-6.6566&3 &191.3065&83 &74003&304&41 04141  & 40 04041  \\
  286644.4894 &-3.4418&3 &191.3065&81 &74003&304&41 04140  & 40 04039  \\
  286644.4894 &-3.4312&3 &191.3065&83 &74003&304&41 04141  & 40 04040  \\
  286644.4894 &-3.4206&3 &191.3065&85 &74003&304&41 04142  & 40 04041  \\
  286644.5243 &-6.6566&3 &191.3065&81 &74003&304&41 04140  & 40 04040  \\
  293624.2500 &-6.6663&3 &200.8680&85 &74003&304&42 04242  & 41 04142  \\
  293624.2841 &-3.4304&3 &200.8680&83 &74003&304&42 04241  & 41 04140  \\
  293624.2841 &-3.4200&3 &200.8680&85 &74003&304&42 04242  & 41 04141  \\
  293624.2842 &-3.4097&3 &200.8680&87 &74003&304&42 04243  & 41 04142  \\
  293624.3191 &-6.6663&3 &200.8680&83 &74003&304&42 04241  & 41 04141  \\
\hline
\multicolumn{9}{|c|}{$^a$ Calculated frequency in MHz}\\
\multicolumn{9}{|c|}{in units of MHz then uncertainty of the line is greater or equal to zero.}\\
\multicolumn{9}{|c|}{$^c$ Base 10 logarithm of the integrated intensity at 300K in nm$^2$ MHz}\\
\multicolumn{9}{|c|}{$^d$ Degrees of freedom in the rotational partition function} \\
\multicolumn{9}{|c|}{(0 for atoms, 2 for linear molecules, 3 for non linear molecules)}\\
\multicolumn{9}{|c|}{$^e$ Lower state energy in cm$^{-1}$ relative to the lowest energy} \\
\multicolumn{9}{|c|}{level in the ground vibrionic state.}\\
\multicolumn{9}{|c|}{$^f$ Upper state degeneracy : g$_{up}=g_{I} \times g_{N}$, where g$_{I}$} \\
\multicolumn{9}{|c|}{is the spin statistical weight and g$_{N} =2N+1$ the rotational degeneracy.}\\
\multicolumn{9}{|c|}{$^g$ Molecule Tag}\\
\multicolumn{9}{|c|}{$^h$ Coding for the format of quantum numbers.}\\
\multicolumn{9}{|c|}{QnF=$100 \times Q + 10 \times H + N_{Qn}$; N$_{Qn}$ is the number of quantum }\\
\multicolumn{9}{|c|}{numbers for each state; H indicates the number of half integer quantum numbers;}\\
\multicolumn{9}{|c|}{Qmod5, the residual when Q is divided by 5, gives the number of principal}\\
\multicolumn{9}{|c|}{quantum numbers (without the spin designating ones)}\\
\multicolumn{9}{|c|}{$^i$ Quantum numbers for the upper state}\\
\multicolumn{9}{|c|}{$^j$ Quantum numbers for the lower state}\\
\hline
\end{tabular}}

\end{document}